\documentclass{aa}
\usepackage[varg]{txfonts}
\usepackage{graphicx}
\usepackage{subcaption}

\begin{document}

    \title{Influence of the interstellar magnetic field
    and 11-year cycle of solar activity
    on the heliopause nose location}
    \titlerunning{Influence of interstellar magnetic field
    on heliopause nose}
    \authorrunning{Bladek, P. \& Ratkiewicz, R.}
    \author{Piotr Bladek \and Romana Ratkiewicz}
    \institute{Space Research Centre Polish Academy of Sciences,
        Bartycka 18A, 00-716 Warsaw, Poland \\
        \email{pbladek@cbk.waw.pl}}
    \abstract{The heliosphere is formed by the interaction between the solar wind (SW) plasma
    emanating from the Sun and a magnetised component of local interstellar medium (LISM) inflowing
    on the Sun. A separation surface called the heliopause (HP) forms between the SW and the LISM.}
    {In this article, we define the nose of the HP and investigate
    the variations in its location. These result from a dependence on the intensity and direction
    of the interstellar magnetic field (ISMF), which is still not well known but has a significant
    impact on the movement of the HP nose, as we try to demonstrate in this paper.}
    {We used a parametric study method based on numerical simulations of various
    forms of the heliosphere using a time-dependent three-dimensional
    magnetohydrodynamic (3D MHD) model of the heliosphere.}
    {The results confirm that the nose of the HP is always in a direction that is perpendicular
    to the maximum ISMF intensity directly behind the HP. The displacement of the HP nose depends
    on the direction and intensity of the ISMF, with the structure of the heliosphere and the
    shape of the HP depending on the 11-year cycle of solar activity.}
    {In the context of the planned space mission to send the Interstellar Probe (IP) to a distance of 1000 AU from the Sun, our study may
    shed light on the question as to which direction the IP should be sent. Further research is needed that introduces elements such as current sheet, reconnection, cosmic
    rays, instability, or turbulence into the models.}
    \keywords{sun: heliosphere -- interplanetary medium -- magnetohydrodynamics (MHD) --
    sun: magnetic fields -- sun: solar wind -- ISM: magnetic fields}
    \maketitle

    \section{Introduction}
    \label{sec:Introduction}

\begin{figure*}
        \sidecaption
        \includegraphics[width=12cm]{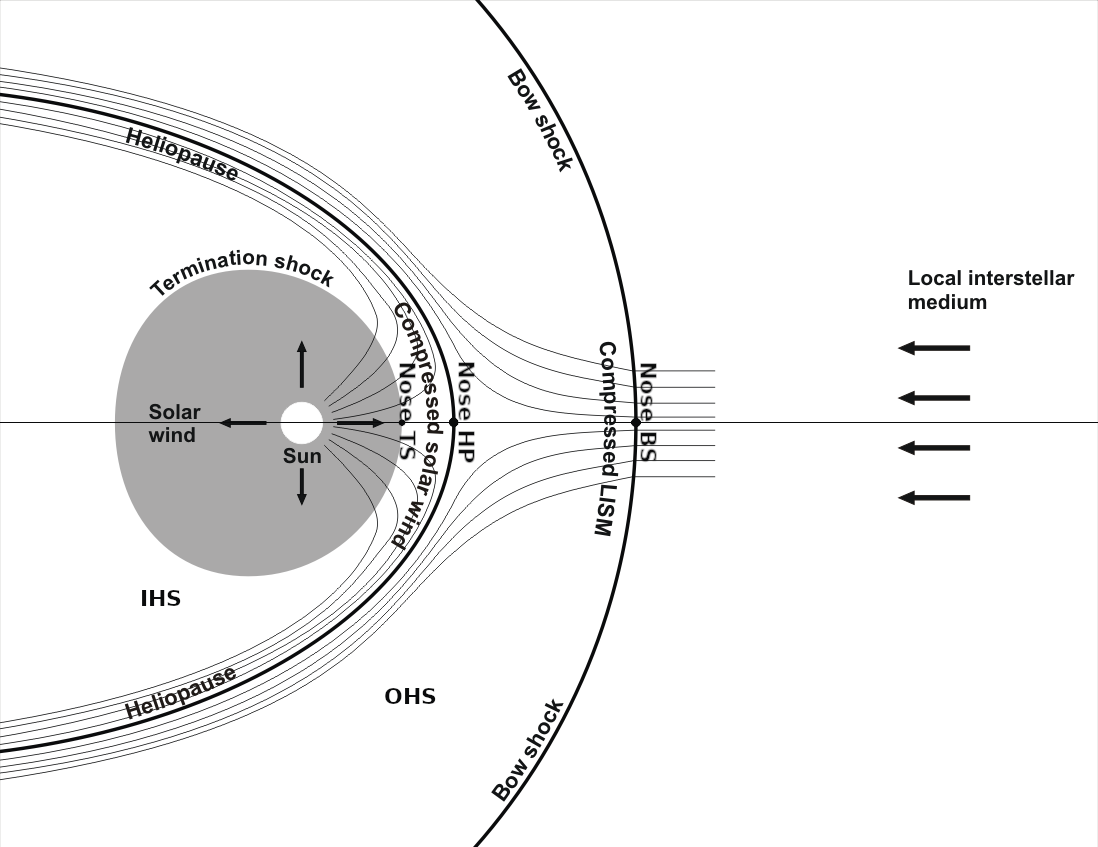}
        \caption{Schematic of the IHS and OHS
        in the parallel LISM velocity and magnetic field vectors (axisymmetric case).
        The TS, HP, and BS noses are on one line.}
        \label{fig:SchematicAxisimetricCase}
    \end{figure*}

    The interaction of the supersonic solar wind (SW) with the local
    interstellar medium (LISM) leads to the formation of a cavity in the LISM
    called the heliosphere, which is filled with the SW plasma. The
    interaction between the SW plasma outflowing spherically and symmetrically
    from the Sun and the counter-flowing LISM plasma in a uniform
    rectilinear motion, and neglecting the magnetic fields in both
    media, results in the axisymmetric shape of the heliosphere. The axial
    symmetry of the heliosphere also holds when the direction of the
    interstellar magnetic field (ISMF) is the same as that of the LISM inflow,
    if the interplanetary magnetic field (IMF) is neglected.
    Mathematically speaking, the heliosphere is separated from the LISM by the heliopause (HP), a
    discontinuity surface that is the pressure
    equilibrium surface of both media. The supersonic SW slows down before
    the HP through a shock wave called the termination shock (TS). If
    the interstellar plasma is also supersonic, it slows down on the other
    side of HP through a shock wave known as a bow shock (BS). The area
    between the TS and the HP is an inner heliosheath (IHS). A layer
    located between the HP and the BS is an outer heliosheath (OHS),
    if the BS exists. Otherwise, the OHP between the HP and the region where
    the LISM is disturbed by the heliosphere causes flowing around the HP.
    In this way, the SW plasma flows around the inner part of the HP,
    and the LISM plasma flows around the outer part of the HP
    (Fig.~\ref{fig:SchematicAxisimetricCase}).
    In an axisymmetric heliosphere, a line running through the
    centre of the Sun and parallel to the velocity vector of the LISM
    intersects the surfaces of the TS, HP, and BS at points that are
    called the noses of the TS, HP, and BS (Fig.~\ref{fig:SchematicAxisimetricCase}).

    \begin{figure*}
        \sidecaption
        \includegraphics[width=12cm]{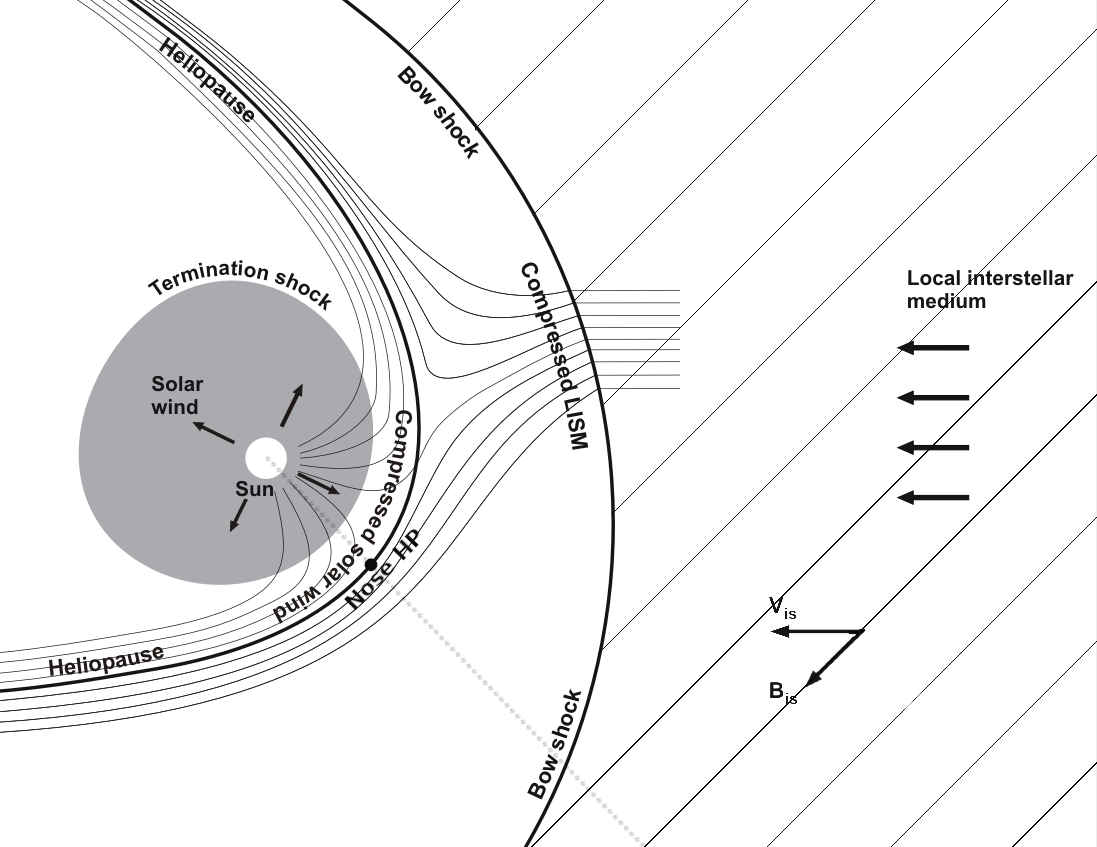}
        \caption{Schematic of the heliosphere and LISM for an ISMF not parallel to the LISM velocity vector. The TS and HP noses
        deviate in one direction (quasi-perpendicular to the
        undisturbed $\pmb{B_{is}}$), and the BS nose in the opposite
        direction \citep{ratkiewicz2000}.}
        \label{fig:SchematicObliqueCase}
    \end{figure*}

    When we include the ISMF $\pmb{B_{is},}$ not parallel to the LISM
    velocity vector $\pmb{V_{is}}$, the noses of TS, HP, and BS change positions
    (Fig.~\ref{fig:SchematicObliqueCase}). The noses of the TS and HP deflect in one direction and
    the nose of the BS deflects in the other direction (Fig.~\ref{fig:SchematicObliqueCase})
    \citep[compare with][]{ratkiewicz1998, ratkiewicz2000}.

    The nose of the heliosphere is cited in various contexts in many publications.
    In some, the nose is identified with the stagnation point of the
    interstellar medium flow
    (e.g. \citet[][]{drake2010, desai2015, dayeh2019, mccomas2020} and \citet{shrestha2023}).
    Many articles refer to the HP nose as 'the nose region of the heliosheath',
    'the nose of the heliosphere' (e.g. \citet[][]{mccomasschwadron2006, lee2009, fiskgloeckler2009,
        fiskgloeckler2014, fiskgloeckler2015, galli2019, kornbleuth2020, kornbleuth2021a,
        opher2017} and \citet{shrestha2023}), 'the nose direction'
    (e.g. \citet[][]{muller2008, opher2013, opher2015, opher2017, opher2021} and \citet{zirnstein2020}) or
    'the upwind (nose) direction' (e.g. \citet[][]{mccomas2020} and \citet[][]{kornbleuth2021b}).
    However, relatively little attention has been paid to the consideration
    of the location of the heliosphere nose, more precisely the location of
    the noses of the TS, HP, and BS and the displacement of the nose.
    This issue was first addressed in the 1980s, using the so-called Newtonian approximation as a model of
    the heliosphere (see Fig. 2 and Fig. 3 \citet{fahr1986}, \citet{fahr1988},
    \citet{ratkiewiczbanaszkiewicz1987} and \citet{banaszkiewiczratkiewicz1989}).
    Recently, this issue has been revisited in \citep{ratkiewiczbaraniecka2023},
    where the HP nose is defined in Fig.~\ref{fig:SchematicObliqueCase}.
    The above articles show that the location of the HP nose depends
    upon the direction and intensity of the ISMF
    and always deflects in a direction quasi-perpendicular to the direction
    of the undisturbed ISMF lines \citep{ratkiewicz2000}.

\begin{figure*}
        \centering
        \includegraphics[width=17cm]{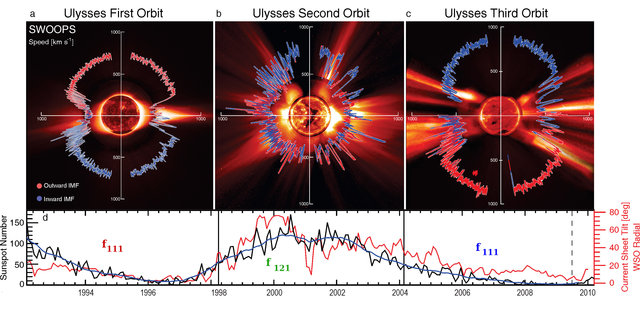}
        \caption{Adapted Fig. 1 (a-d) from \citet{mccomas2008}, original caption
        '(a–c) Polar plots of the solar wind speed, colored by IMF polarity for Ulysses`
        three polar orbits colored toindicate measured magnetic polarity.
        In each,the earliest times are on the left (nine o`clock position) and progress
        aroundcounterclockwise. (d) Contemporaneous values for the smoothed sunspot number (black) and
        heliospheric current sheettilt (red), lined up to match Figures 1a–1c. In Figures 1a–1c, the
        solar wind speed is plotted over characteristic solarimages for solar minimum for
        cycle 22 (8/17/96), solar maximum for cycle 23 (12/07/00), and solar minimum for cycle 23(03/28/06).
        From the center out, we blend images from the Solar and
        Heliospheric Observatory (SOHO) Extremeultraviolet Imaging Telescope (Fe XII at 1950 nm), the Mauna Loa K
        coronameter (700–950 nm), and the SOHO C2white light coronagraph'.}
        \label{fig:McComas2008}
    \end{figure*}

    In this article, using the definition of the HP nose in Fig.~\ref{fig:SchematicAxisimetricCase}
    \citep[compare with Fig. 2 in][]{ratkiewiczbaraniecka2023}, we discuss possible
    configurations of the HP nose, arising under the influence of various
    ISMF intensities and directions, for the SW velocity during the minimum
    and maximum of the 11-year cycle of solar activity, (Fig.~\ref{fig:McComas2008}).
    We show that the HP nose clearly deviates from the
    LISM inflow direction and is always directed towards the ISMF maximum,
    just behind the HP.

    The paper is organised as follows: in Sect.~\ref{sec:SimulationMethodAndBoundaryConditions},
    we describe the simulation
    method and three sets of boundary conditions; in Sect.~\ref{sec:Results}, we present
    the results of our modelling; in Sect.\ 4, we present our conclusions.

    \section{Simulation method and boundary conditions}
    \label{sec:SimulationMethodAndBoundaryConditions}

    We used a three-dimensional (3D) magnetohydrodynamic (MHD) model
    of the interaction between the SW and LISM,
    as described by \citet{ratkiewicz2002, ratkiewicz2008},
    with a number of revisions introduced, including:
    a magnetised SW, the boundary conditions adjusted to simulate solar cycle effects,
    and an improved modelling of the neutral hydrogen within the constant flux approximation,
    as described by \citet{strumikratkiewicz2022}.
    The set of MHD equations in Eq. 1 includes a source term $S$ on the right-hand side
    to describe a resonance charge exchange with a constant flux
    of hydrogen, and a second source term, $Q$, that maintains
    the divergence-free magnetic field \citep{ratkiewicz2002, ratkiewicz2008}.

    \begin{equation}
        \label{eq:mhdEquation}
        \frac{\partial \pmb{U}}{\partial t} + \nabla \cdot \hat{\pmb{F}} = \pmb{Q} + \pmb{S}
    \end{equation},

    where $\pmb{U}$, $\pmb{Q}$, and $\pmb{S}$ are column vectors, and $\hat{\pmb{F}}$ is a flux tensor defined as

    \begin{equation*}
        \begin{array}{c}
            \pmb{U} = \left[ \begin{array}{c}
                                 \rho \\ \rho \pmb{u} \\ \pmb{B} \\ \rho E
            \end{array} \right],
            \hat{\pmb{F}} = \left[ \begin{array}{c}
                                       \rho \pmb{u} \\ \rho \pmb{uu} + \pmb{I}(p + \frac{\pmb{B \cdot B}}{8 \pi}) - \frac{\pmb{BB}}{4\pi} \\ \pmb{uB} - \pmb{Bu} \\ \rho H\pmb{u} - \frac{\pmb{B}(\pmb{u} \cdot \pmb{B})}{4\pi}
            \end{array} \right],
            \\
            \\
            \pmb{Q} = - \left[ \begin{array}{c}
                                   0 \\ \frac{\pmb{B}}{4\pi} \\ \pmb{u} \\ \pmb{u} \cdot \frac{\pmb{B}}{4\pi}
            \end{array} \right] \nabla \cdot \pmb{B},
            \pmb{S} = \rho \nu_{c} \left[ \begin{array}{c}
                                              0 \\ \pmb{V_{H}} - \pmb{u} \\ 0 \\ \frac{1}{2} V_{H}^2 + \frac{3k_{B}T_{H}}{2m_{H}} - \frac{1}{2}u^2-\frac{k_{B}T}{(\gamma - 1)m_{H}}
            \end{array} \right]
        \end{array}
    \end{equation*}\begin{flushright}
                       .
    \end{flushright}

    Here, $\rho$ is the ion mass density, $p = 2nk_{B}T$ is the pressure, $n$ is the ion number
    density, $T$ and $T_H$ ($T_H = const.$) are ion and hydrogen atom temperatures, $\pmb{u}$ and
    $\pmb{V_H}$ ($\pmb{V_H} = const.$) are the ion and hydrogen atom velocity vectors, respectively,
    $\pmb{B}$ is the magnetic field vector, $E = \frac{1}{\gamma - 1} \frac{p}{\rho} + \frac {\pmb{u \cdot u}}{2} + \frac{\pmb{B \cdot B}}{8\pi \rho}$
    is the total energy per unit mass, and $H = \frac{\gamma}{\gamma  - 1} \frac{p}{\rho} + \frac {\pmb{u \cdot u}}{2} + \frac{\pmb{B \cdot B}}{4\pi \rho}$
    is enthalpy. The $\gamma$ is the ratio of specific heats and
    $\pmb{I}$ is the 3 x 3 identity matrix. The charge exchange collision frequency is
    $\nu_c = n_H \sigma u_{s}$, where $n_H$ ($n_H$ = const) is the hydrogen atom number density,
    $\sigma$ is the charge exchange cross-section, and $u_{s} = \sqrt{(\pmb{u} - \pmb{V_{H}})^2 + \frac{128k_B(T + T_H)}{9\pi m_H}}$
    is the effective average relative speed of protons and hydrogen atoms, assuming a Maxwellian
    spread of velocities both for protons and hydrogen atoms.
    The flows are taken to be adiabatic, with $\gamma = \frac{5}{3}$.
    The additional constraint of a divergence-free magnetic
    field, $\nabla \cdot \pmb{B} = 0$, in the numerical simulations is accomplished by adding the
    source term $\pmb{Q}$ to the right-hand side of (Eq.~1), which is proportional to the divergence of the magnetic field.
    Adding $\pmb{Q}$ to the right-hand side of (Eq.~1) assures that any numerically
    generated $\nabla \cdot \pmb{B} \neq 0$ is advected with the flow, and allows one to limit the growth of $\nabla \cdot \pmb{B} \neq 0$.

    In order to thoroughly analyse the movement of the nose HP, we considered three cases:
    \begin{description}
        \item[Case one] 'iso without IMF' - numerical simulations are carried out during the period
        of a maximum of the solar cycle activity for isotropic SW without the
        IMF.
        \item[Case two] 'iso' -  numerical simulations are carried out during the period
        of a maximum of the solar cycle activity for isotropic SW with the IMF
        (see Fig. \ref{fig:McComas2008}b).
        \item[Case three] 'anis' - numerical simulations are carried out during the period
        of a minimum of the solar cycle activity for slow and fast solar wind,
        taking into account the IMF (see Fig. \ref{fig:McComas2008}a and \ref{fig:McComas2008}c).
    \end{description}
    The LISM parameters are set at the outer boundary, $5000 AU$ from the Sun:
    velocities and temperatures of ionised and neutral LISM components are equal
    and $V_{is} = 26.4 km/s$, $T_{is} = 6400 K$, proton number density
    $n_p = 0.06 cm^{-3}$, and neutral hydrogen number density $n_H = 0.11 cm^{-3}$.

    The SW parameters are set at the inner boundary, $10 AU$ from the Sun:
    In case one, velocity $V_{sw} = 420 km/s$, number density SW protons
    $n_p = 0.052 cm^{-3}$, and the IMF is neglected.
    In case two, $V_{sw}$ and $n_{sw}$ are the same as case one, but the IMF is set according
    to the Parker model as an Archimedean spiral, and its strength at $1 AU$ equals 35.5 $\mu G$.
    In case three, for the slow SW, velocity $V_{sw} = 420 km/s$, number density
    $n_p = 0.052 cm^{-3}$; for the fast SW, velocity $V_{sw} = 798 km/s$,
    number density $n_p = 0.027 cm^{-3}$; the IMF for the slow
    and fast SW is the same as case two.

    To show the effect of the ISMF intensity and direction on the deviation
    of the HP nose from the direction of the LISM inflow,
    we considered different HP configurations for the ISMF intensities of
    $2 \mu G$, $3 \mu G$, and $4 \mu G$ and for the angle between the LISM
    velocity vector and the direction of the ISMF vector,
    called an inclination angle $\alpha$,  $0\degr$, $30\degr$, $60\degr$, and $90\degr$.

    \section{Results}
    \label{sec:Results}

    \subsection{The behaviour of the HP in the case of LISM velocity parallel to the ISMF}

\begin{figure}
        \centering
        \begin{subfigure}[b]{0.4\textwidth}
            \centering
            \resizebox{\hsize}{!}{\includegraphics{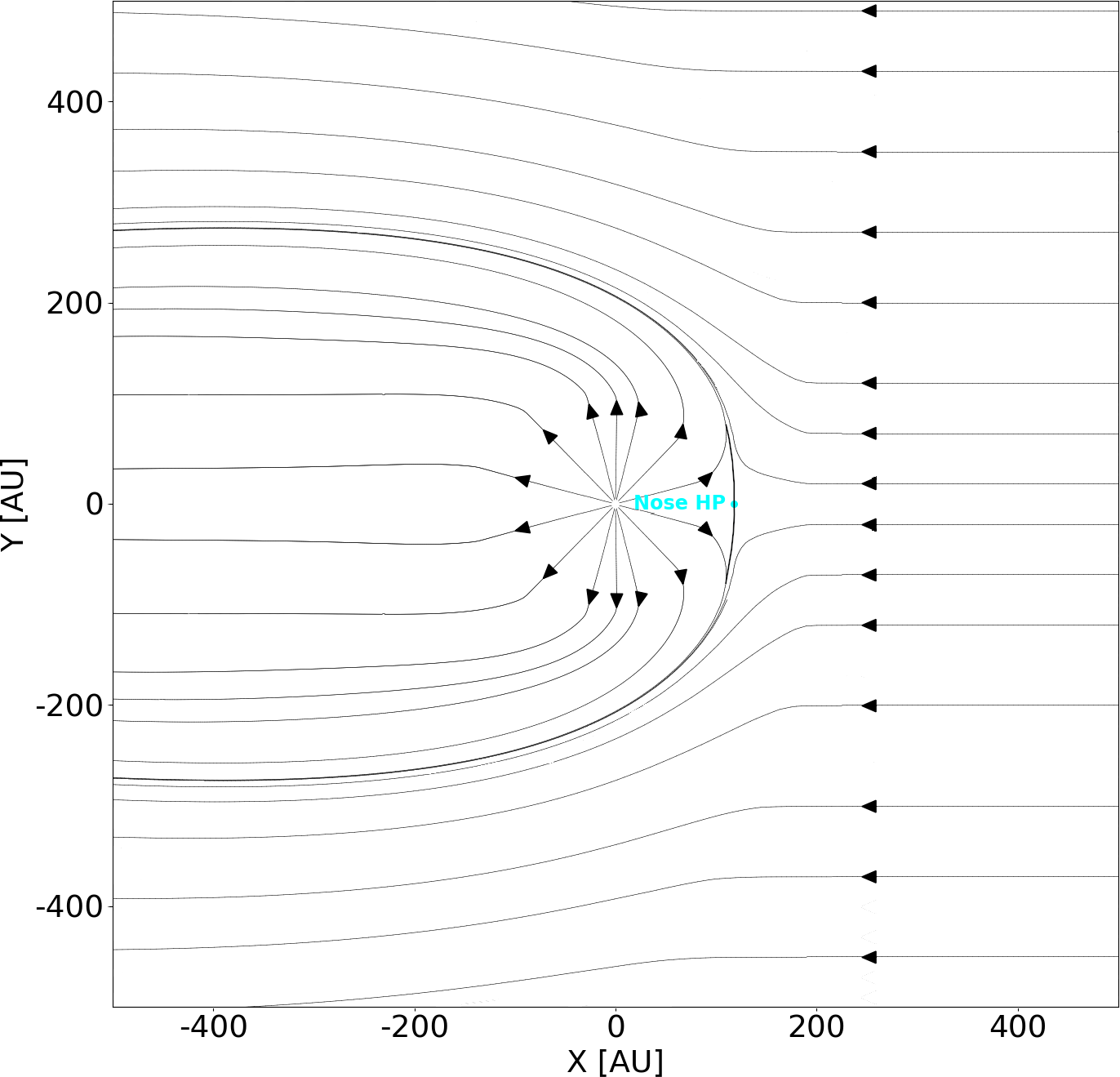}}
            \caption{ISMF intensity equal to $2 \mu G$}
            \label{fig:streamlines2}
        \end{subfigure}
        \vskip\baselineskip
        \begin{subfigure}[b]{0.4\textwidth}
            \centering
            \resizebox{\hsize}{!}{\includegraphics{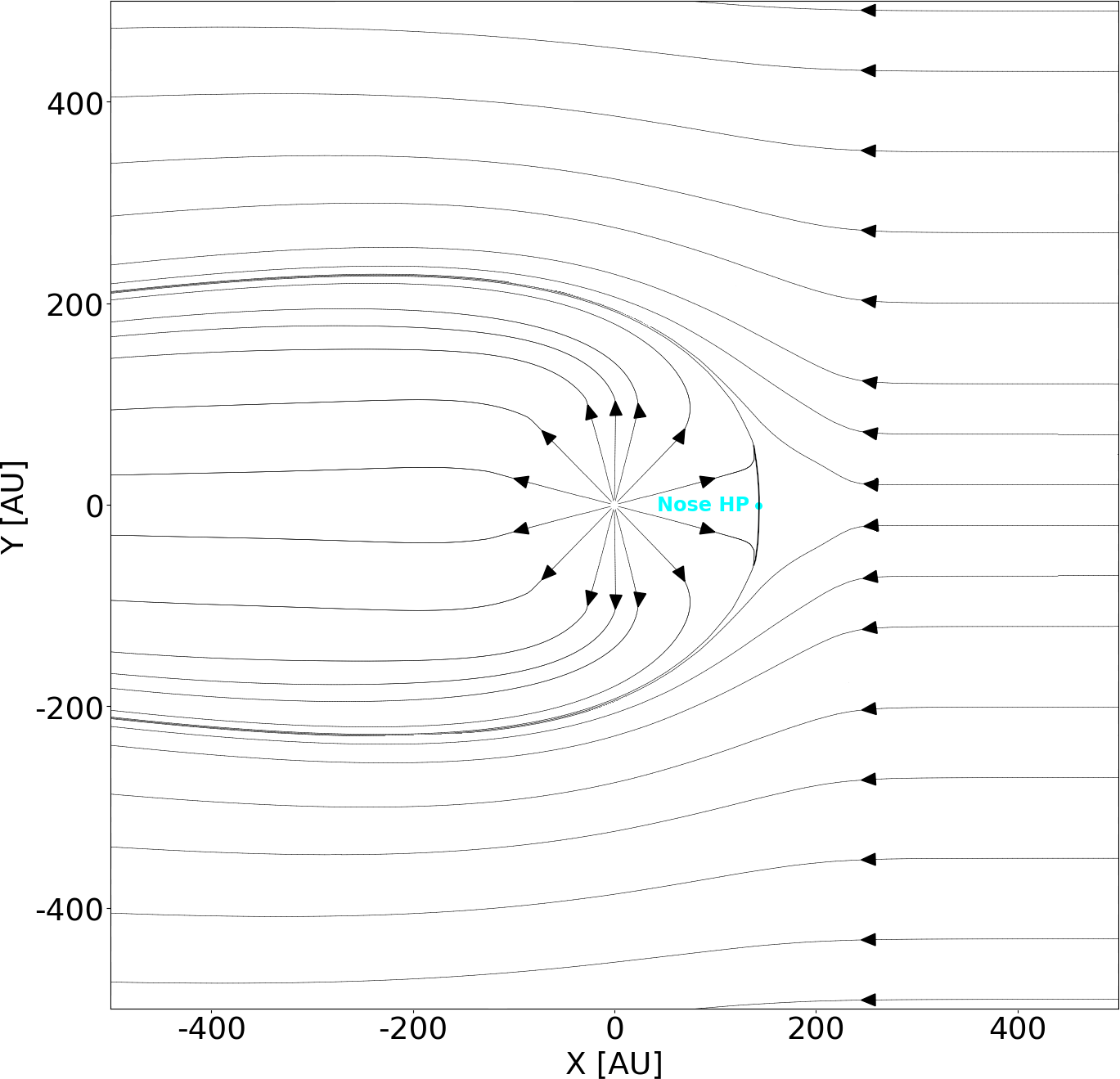}}
            \caption{ISMF intensity equal to $3 \mu G$}
            \label{fig:streamlines3}
        \end{subfigure}
        \vskip\baselineskip
        \begin{subfigure}[b]{0.4\textwidth}
            \centering
            \resizebox{\hsize}{!}{\includegraphics{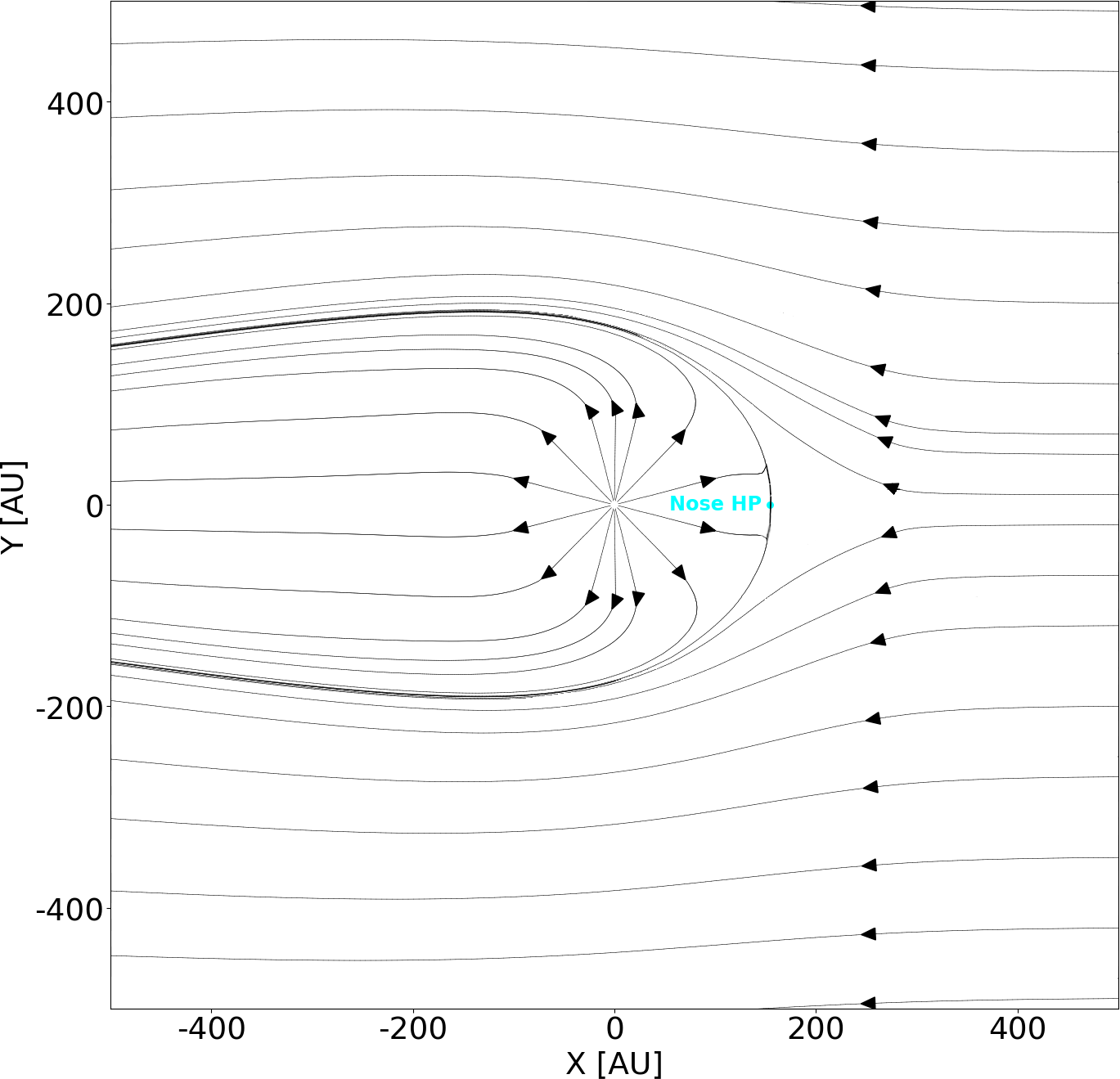}}
            \caption{ISMF intensity equal to $4 \mu G$}
            \label{fig:streamlines4}
        \end{subfigure}
        \caption{HP shape and location of the noses of the HP.
        Velocity streamlines for inclination angle equal $0\degr$
            are shown in the x-y plane}
        \label{fig:streamlines}
    \end{figure}

    The HP shape and HP nose for the three cases of intensity
    of the ISMF with the ISMF direction
    parallel to the LISM velocity vector are shown in Figs.
    \ref{fig:streamlines}\subref{fig:streamlines2},
    \ref{fig:streamlines}\subref{fig:streamlines3},
    and \ref{fig:streamlines}\subref{fig:streamlines4}.

    For the ISMF parallel to the LISM velocity, the heliosphere is axisymmetric,
    so in the x-y and x-z planes it looks the same and in the y-z plane,
    the isolines form circles with the Sun in the centre
    \citep[compare with Figs. 3a, 3b, 3c][]{ratkiewicz2000}.
    As shown in Figs.
    \ref{fig:streamlines}\subref{fig:streamlines2},
    \ref{fig:streamlines}\subref{fig:streamlines3},
    and \ref{fig:streamlines}\subref{fig:streamlines4},
    the characteristic feature for various ISMF intensities is
    that the more the ISMF compresses the
    HP, the greater the ISMF intensity is.
    The HP nose is in each case located at the intersection
    of the x-axis with the HP (blue dots).
    On the other hand, the greater the ISMF intensity,
    the farther the HP nose is from the Sun, so that, in the case of an ISMF parallel to
    the LISM velocity vector, the greater the
    ISMF intensity, the farther the HP nose is extended
    towards the direction of the interstellar medium.

    \begin{figure}
        \centering
        \begin{subfigure}[b]{0.465\textwidth}
            \centering
            \resizebox{\hsize}{!}{\includegraphics{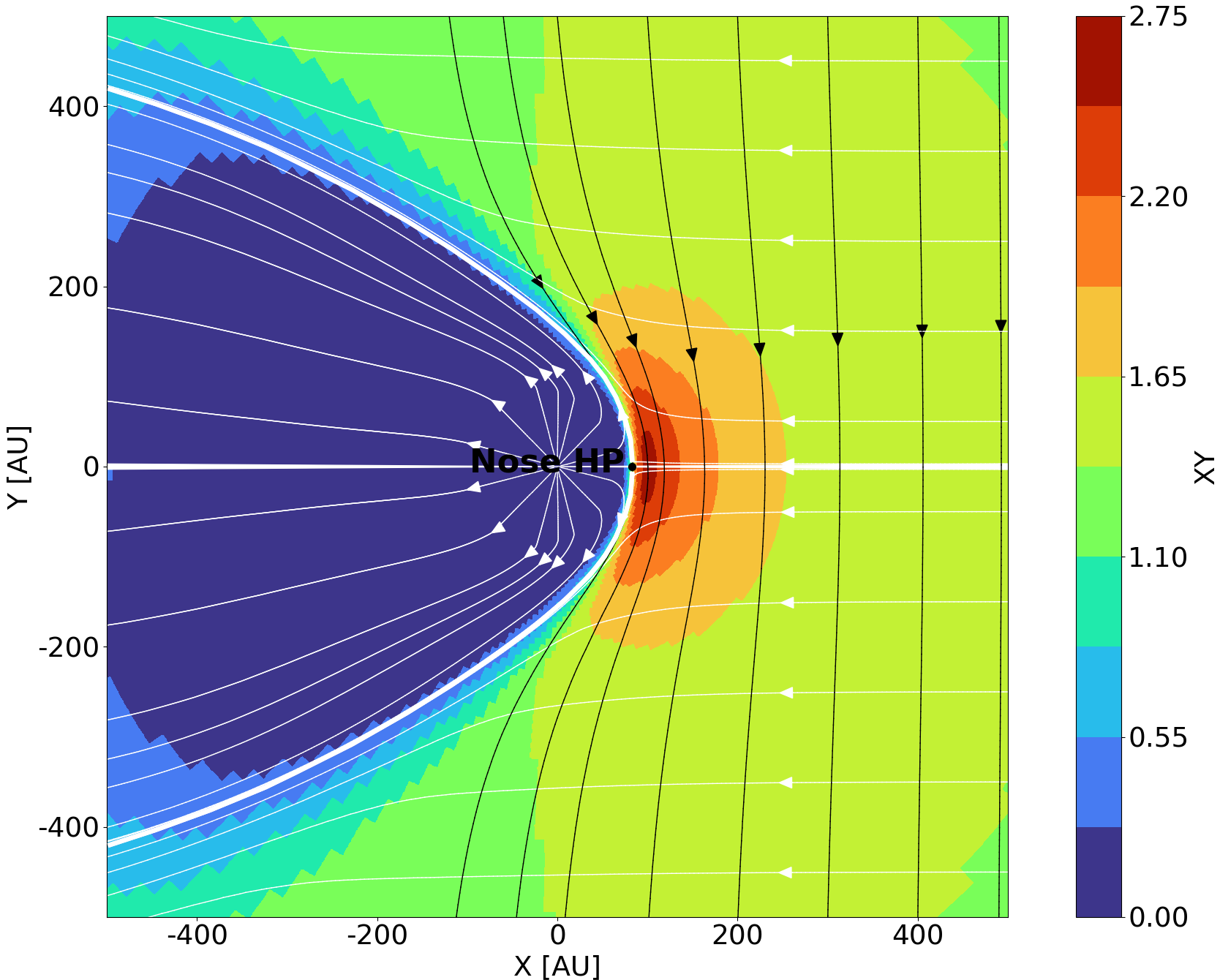}}
            \caption{x-y plane}
            \label{fig:streamlines_fieldlines_4_90_xy}
        \end{subfigure}
        \vskip\baselineskip
        \begin{subfigure}[b]{0.465\textwidth}
            \centering
            \resizebox{\hsize}{!}{\includegraphics{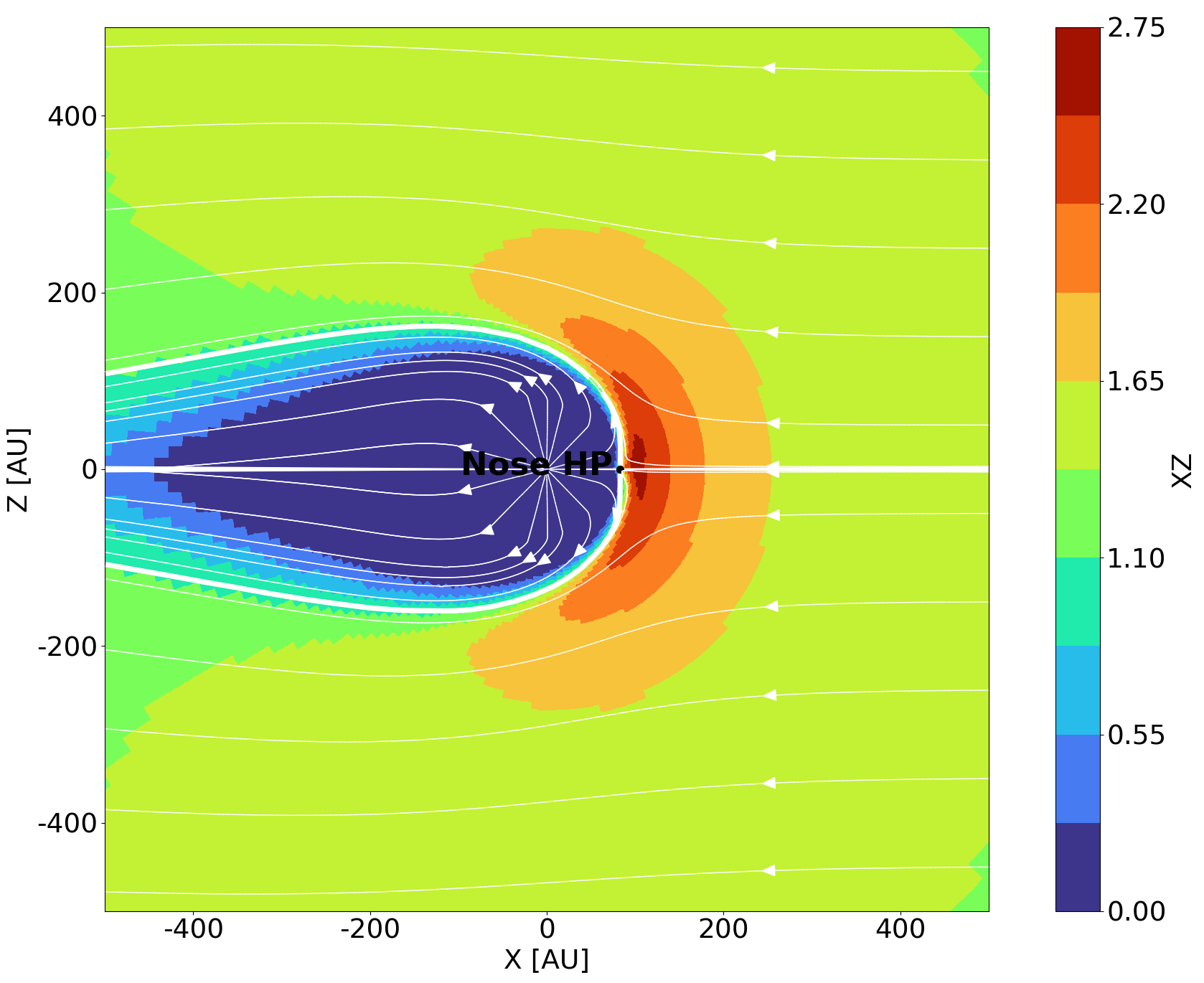}}
            \caption{x-z plane (note, magnetic fieldlines in this case are
            perpendicular to x-z plane and are omitted)}
            \label{fig:streamlines_4_90_xz}
        \end{subfigure}
        \vskip\baselineskip
        \begin{subfigure}[b]{0.465\textwidth}
            \centering
            \resizebox{\hsize}{!}{\includegraphics{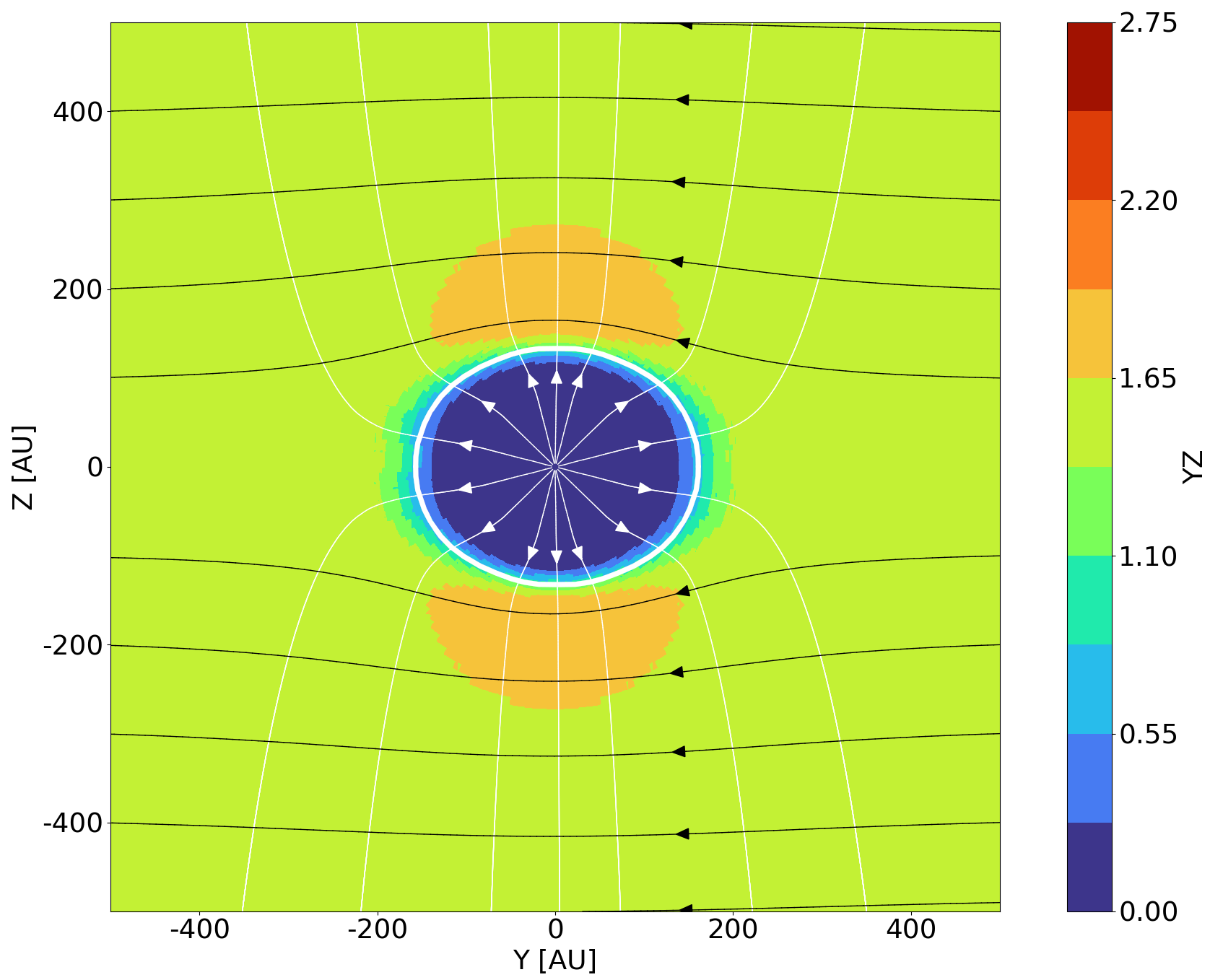}}
            \caption{y-z plane}
            \label{fig:streamlines_fieldlines_4_90_yz}
        \end{subfigure}
        \caption{Velocity streamlines (white) and magnetic fieldlines
            (black) shown for the ISMF magnitude for the inclination
            angle equal to $90\degr$ and for ISMF intensity $4 \mu G$,
            without the IMF}
        \label{fig:streamlines_fieldlines_4_90}
    \end{figure}

    \begin{figure}
        \resizebox{\hsize}{!}{\includegraphics{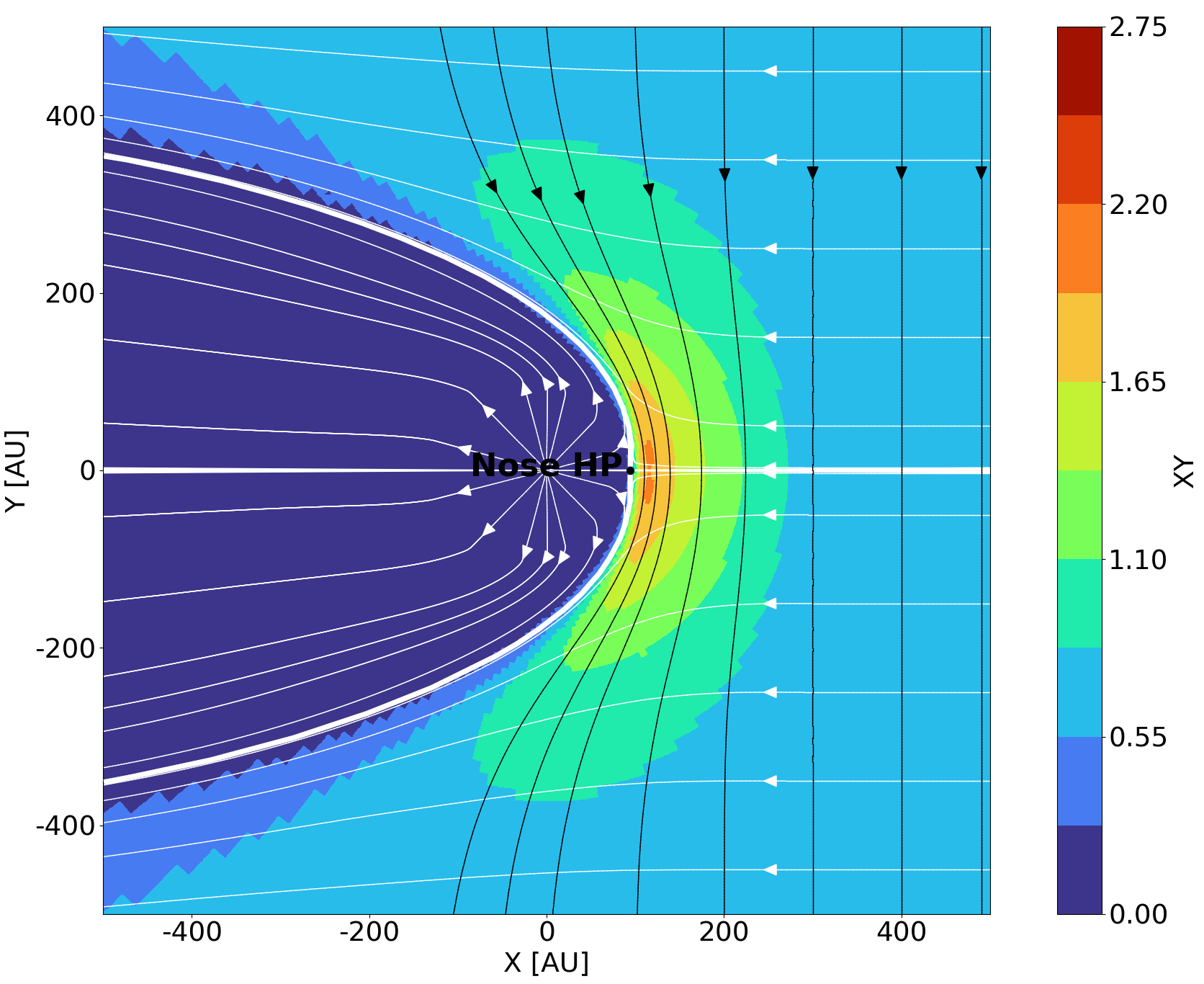}}
        \caption{Velocity streamlines (white) and magnetic fieldlines (black)
            shown for the ISMF magnitude for the inclination angle
            equal to $90\degr$ and for ISMF intensity $2 \mu G$,
            without the IMF, in the x-y plane}
        \label{fig:streamlines_fieldlines_2_90_xy}
    \end{figure}

    \subsection{The behaviour of the HP in the case of LISM velocity perpendicular to the ISMF}

    The heliosphere for case one, iso without IMF, of the
    ISMF direction perpendicular to the LISM velocity
    is shown in the x-y, x-z, and y-z planes in
    Figs
    \ref{fig:streamlines_fieldlines_4_90}\subref{fig:streamlines_fieldlines_4_90_xy},
    \ref{fig:streamlines_fieldlines_4_90}\subref{fig:streamlines_4_90_xz}, and
    \ref{fig:streamlines_fieldlines_4_90}\subref{fig:streamlines_fieldlines_4_90_yz}
    for an ISMF intensity of $4 \mu G$
    to illustrate the shape of the HP and the nose location.

    For an ISMF perpendicular to the LISM velocity,
    the heliosphere (in comparison to the parallel field described above)
    loses its axial symmetry. In the x-y plane, the nose of the
    HP, squeezed by the perpendicular lines of the ISMF field,
    approaches the Sun (see Fig.
    \ref{fig:streamlines_fieldlines_4_90}\subref{fig:streamlines_fieldlines_4_90_xy}),
    and the profile of the HP towards its tail increases its distance
    from the Sun. In the x-z plane, the HP is compressed along
    the z-axis (see Fig.
    \ref{fig:streamlines_fieldlines_4_90}\subref{fig:streamlines_4_90_xz}).
    In the y-z plane (see Fig.
    \ref{fig:streamlines_fieldlines_4_90}\subref{fig:streamlines_fieldlines_4_90_yz}),
    the HP has the shape of a flattened circle
    \citep[compare with Fig. 4][]{ratkiewicz2000}.
    As can be seen in Figs.
    \ref{fig:streamlines_fieldlines_4_90}\subref{fig:streamlines_fieldlines_4_90_xy},
    \ref{fig:streamlines_fieldlines_4_90}\subref{fig:streamlines_4_90_xz},
    and \ref{fig:streamlines_fieldlines_4_90}\subref{fig:streamlines_fieldlines_4_90_yz},
    the maximum ISMF intensity is in the direction perpendicular to
    the Sun's line of sight \citep[compare][]{ratkiewicz2000}.

    Figure \ref{fig:streamlines_fieldlines_2_90_xy} shows the same results as
    Fig. \ref{fig:streamlines_fieldlines_4_90}\subref{fig:streamlines_fieldlines_4_90_xy},
    except for the ISMF intensity, which in
    Fig. \ref{fig:streamlines_fieldlines_4_90}\subref{fig:streamlines_fieldlines_4_90_xy}
    is two times greater ($4 \mu G$) than in
    Fig. \ref{fig:streamlines_fieldlines_2_90_xy} ($2 \mu G$).
    This comparison shows the greater compression of the HP nose in
    Fig. \ref{fig:streamlines_fieldlines_4_90}\subref{fig:streamlines_fieldlines_4_90_xy},
    which is manifested by a shorter distance of the HP nose
    and the Sun and by greater distances between the heliosphere
    surface and the x-axis towards the tail.

    \subsection{The behaviour of the HP in the case of ISMF direction oblique to the LISM velocity}

    In numerical simulations, two inclination angles from the
    range $0\degr< \alpha < 90\degr$ are taken into account,
    namely, $\alpha = 30\degr$ and $\alpha = 60\degr$.
    In order to determine the offset of the HP nose,
    cases one through three (see Sect.~\ref{sec:SimulationMethodAndBoundaryConditions}) were first analysed for the angle $\alpha = 60\degr$.

\begin{figure}
        \centering
        \begin{subfigure}[b]{0.455\textwidth}
            \centering
            \resizebox{\hsize}{!}{\includegraphics{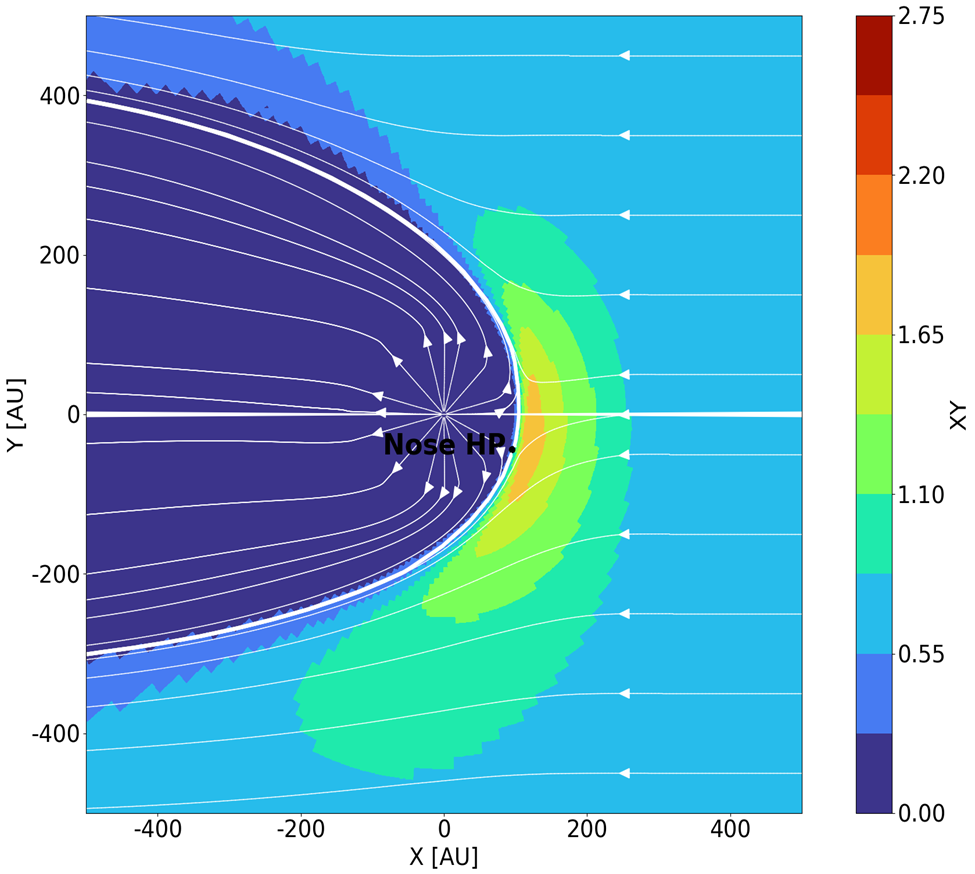}}
            \caption{Iso SW without the IMF}
            \label{fig:streamlines_60_2_iso_no_bsw}
        \end{subfigure}
        \vskip\baselineskip
        \begin{subfigure}[b]{0.455\textwidth}
            \centering
            \resizebox{\hsize}{!}{\includegraphics{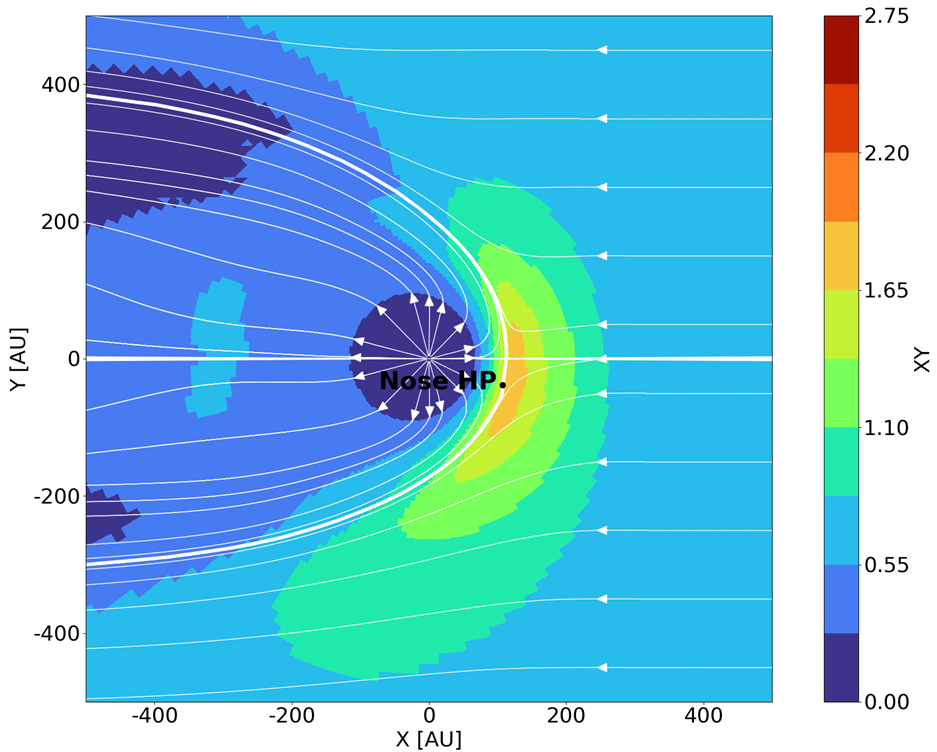}}
            \caption{Iso SW with the IMF}
            \label{fig:streamlines_60_2_iso}
        \end{subfigure}
        \vskip\baselineskip
        \begin{subfigure}[b]{0.455\textwidth}
            \centering
            \resizebox{\hsize}{!}{\includegraphics{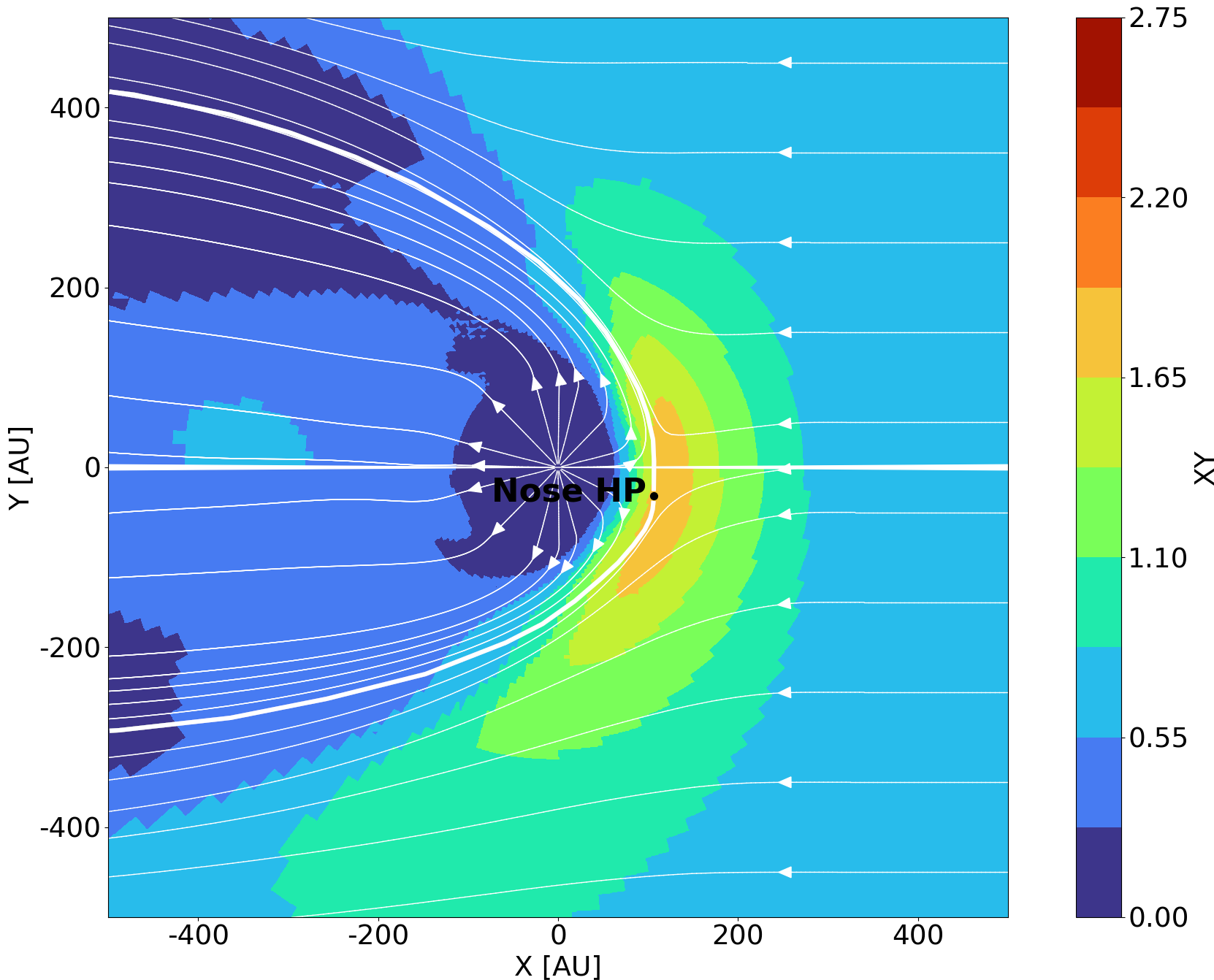}}
            \caption{Anis SW with the IMF}
            \label{fig:streamlines_60_2_anis}
        \end{subfigure}
        \caption{Velocity streamlines (white) shown for the magnetic field
        magnitude for the inclination angle $\alpha = 60\degr$ and for
        ISMF intensity $2 \mu G$}
        \label{fig:streamlines_60_2}
    \end{figure}

\begin{figure}
        \centering
        \begin{subfigure}[b]{0.465\textwidth}
            \centering
            \resizebox{\hsize}{!}{\includegraphics{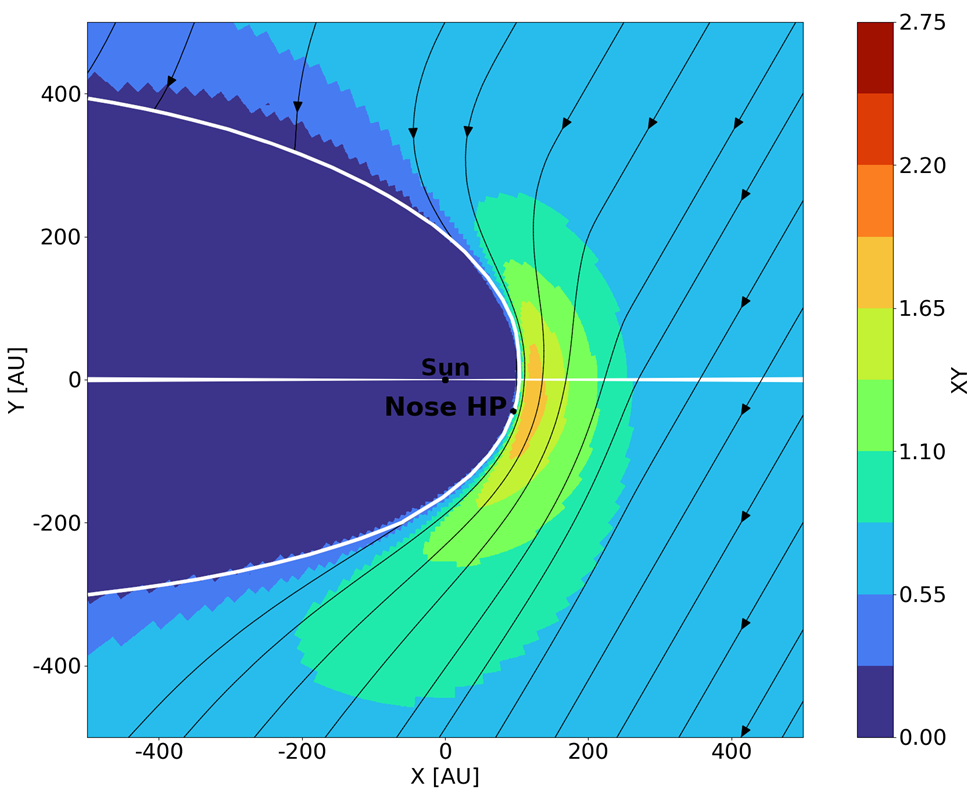}}
            \caption{Iso SW without the IMF (compare to \ref{fig:streamlines_60_2}\subref{fig:streamlines_60_2_iso_no_bsw})}
            \label{fig:fieldlines_60_2_iso_no_bsw}
        \end{subfigure}
        \vskip\baselineskip
        \begin{subfigure}[b]{0.465\textwidth}
            \centering
            \resizebox{\hsize}{!}{\includegraphics{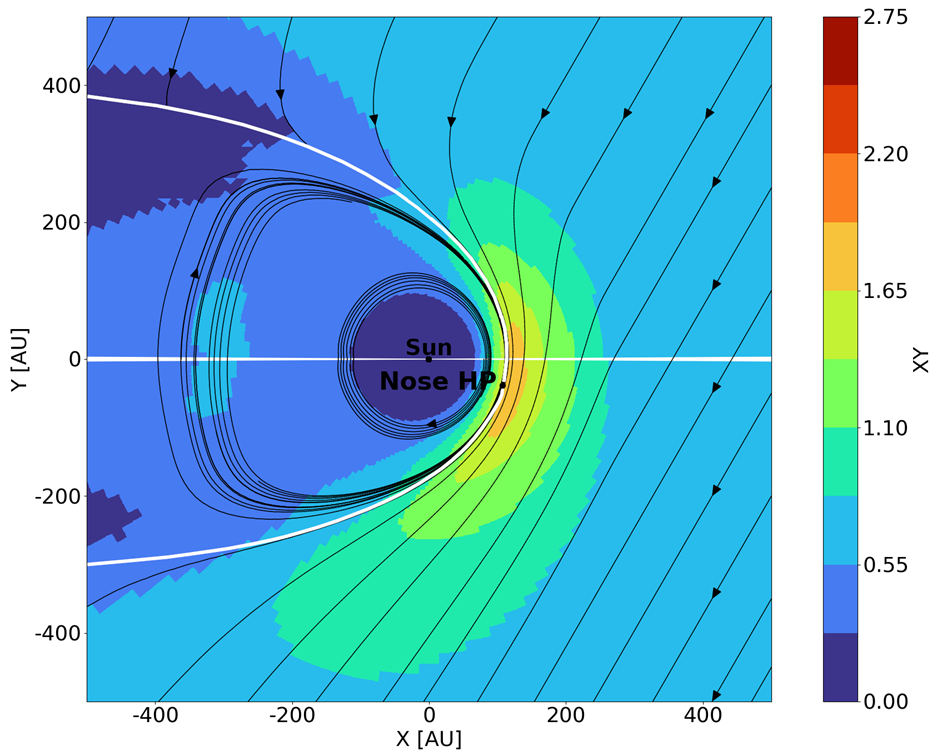}}
            \caption{Iso SW with the IMF (compare to \ref{fig:streamlines_60_2}\subref{fig:streamlines_60_2_iso})}
            \label{fig:fieldlines_60_2_iso}
        \end{subfigure}
        \vskip\baselineskip
        \begin{subfigure}[b]{0.465\textwidth}
            \centering
            \resizebox{\hsize}{!}{\includegraphics{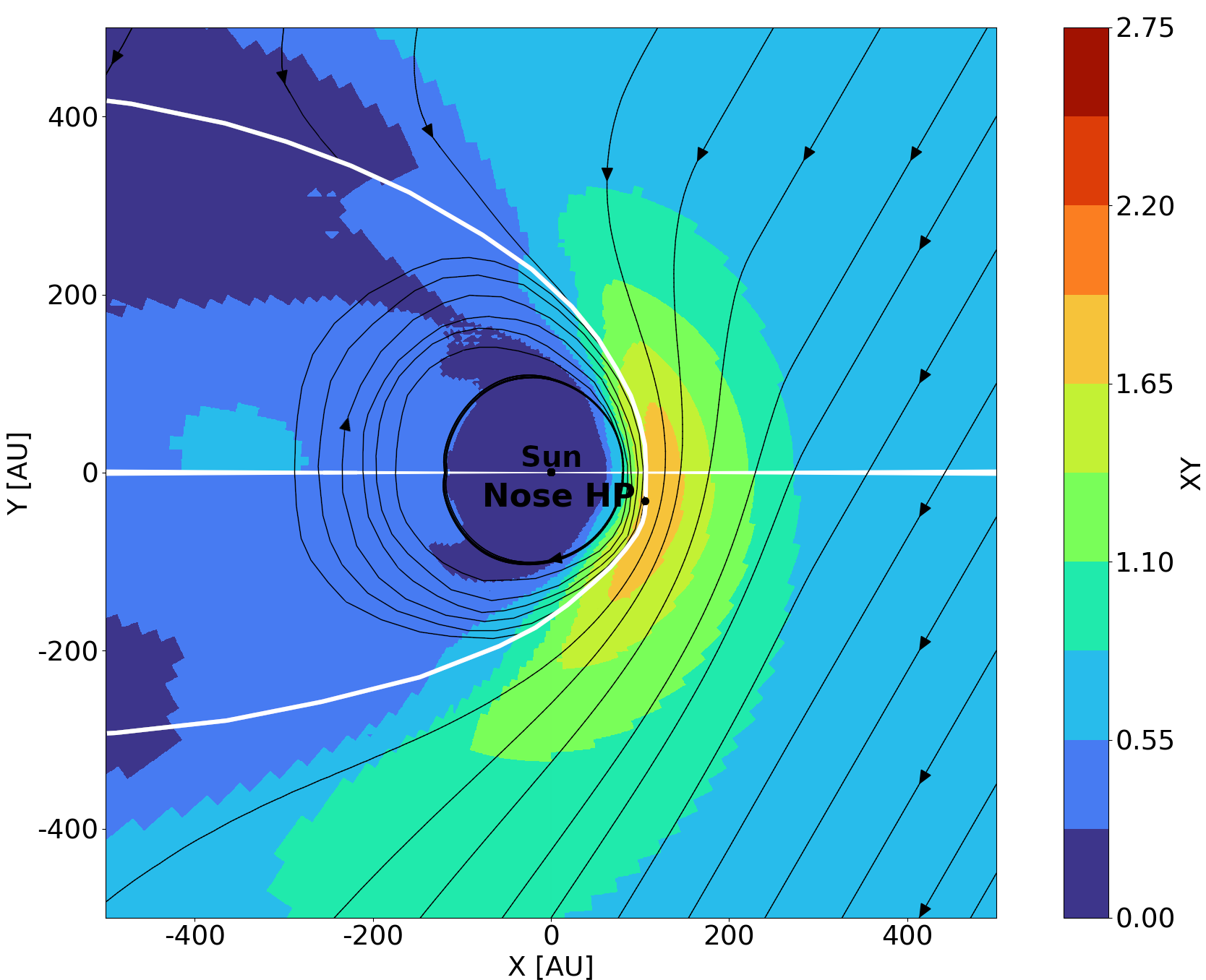}}
            \caption{Anis SW with the IMF (compare to \ref{fig:streamlines_60_2}\subref{fig:streamlines_60_2_anis})}
            \label{fig:fieldlines_60_2_anis}
        \end{subfigure}
        \caption{Magnetic fieldlines (black) shown for the magnetic
        field magnitude for the inclination angle $\alpha = 60\degr$ and
        for ISMF intensity $2 \mu G$}
        \label{fig:fieldlines_60_2}
    \end{figure}

    Figures
    \ref{fig:streamlines_60_2}\subref{fig:streamlines_60_2_iso_no_bsw},
    \ref{fig:streamlines_60_2}\subref{fig:streamlines_60_2_iso},
    and \ref{fig:streamlines_60_2}\subref{fig:streamlines_60_2_anis} show the deviation of the HP nose from the
    direction of the LISM velocity and the shape of the
    HP for the ISMF intensity $2 \mu G$ using the streamlines.
    To better analyse the behaviour of the HP nose,
    comparisons with Figs.
    \ref{fig:streamlines_60_2}\subref{fig:streamlines_60_2_iso_no_bsw},
    \ref{fig:streamlines_60_2}\subref{fig:streamlines_60_2_iso},
    and \ref{fig:streamlines_60_2}\subref{fig:streamlines_60_2_anis}
    and Figs.
    \ref{fig:fieldlines_60_2}\subref{fig:fieldlines_60_2_iso_no_bsw},
    \ref{fig:fieldlines_60_2}\subref{fig:fieldlines_60_2_iso},
    and \ref{fig:fieldlines_60_2}\subref{fig:fieldlines_60_2_anis}
    were created, which show the same results as Figs.
    \ref{fig:streamlines_60_2}\subref{fig:streamlines_60_2_iso_no_bsw},
    \ref{fig:streamlines_60_2}\subref{fig:streamlines_60_2_iso},
    and \ref{fig:streamlines_60_2}\subref{fig:streamlines_60_2_anis},
    but with the use of magnetic fieldlines.
    It is easy to see that the HP for solar maximum,
    without an IMF (case one; see Figs.
    \ref{fig:streamlines_60_2}\subref{fig:streamlines_60_2_iso_no_bsw} and
    \ref{fig:fieldlines_60_2}\subref{fig:fieldlines_60_2_iso_no_bsw})
    is greater than for case two with the IMF (see Figs.
    \ref{fig:streamlines_60_2}\subref{fig:streamlines_60_2_iso} and
    \ref{fig:fieldlines_60_2}\subref{fig:fieldlines_60_2_iso}).
    The heliosphere for the minimum 11-year cycle of solar
    activity, with the IMF taken into account, (case three; see Figs.
    \ref{fig:streamlines_60_2}\subref{fig:streamlines_60_2_anis} and
    \ref{fig:fieldlines_60_2}\subref{fig:fieldlines_60_2_anis})
    differs from the previous examples (Figs.
    \ref{fig:streamlines_60_2}\subref{fig:streamlines_60_2_iso} and
    \ref{fig:streamlines_60_2}\subref{fig:streamlines_60_2_iso}).
    In all three cases, the nose of the HP clearly deviates
    from the direction of the LISM velocity.

    \begin{figure}
        \centering
        \begin{subfigure}[b]{0.465\textwidth}
            \centering
            \resizebox{\hsize}{!}{\includegraphics{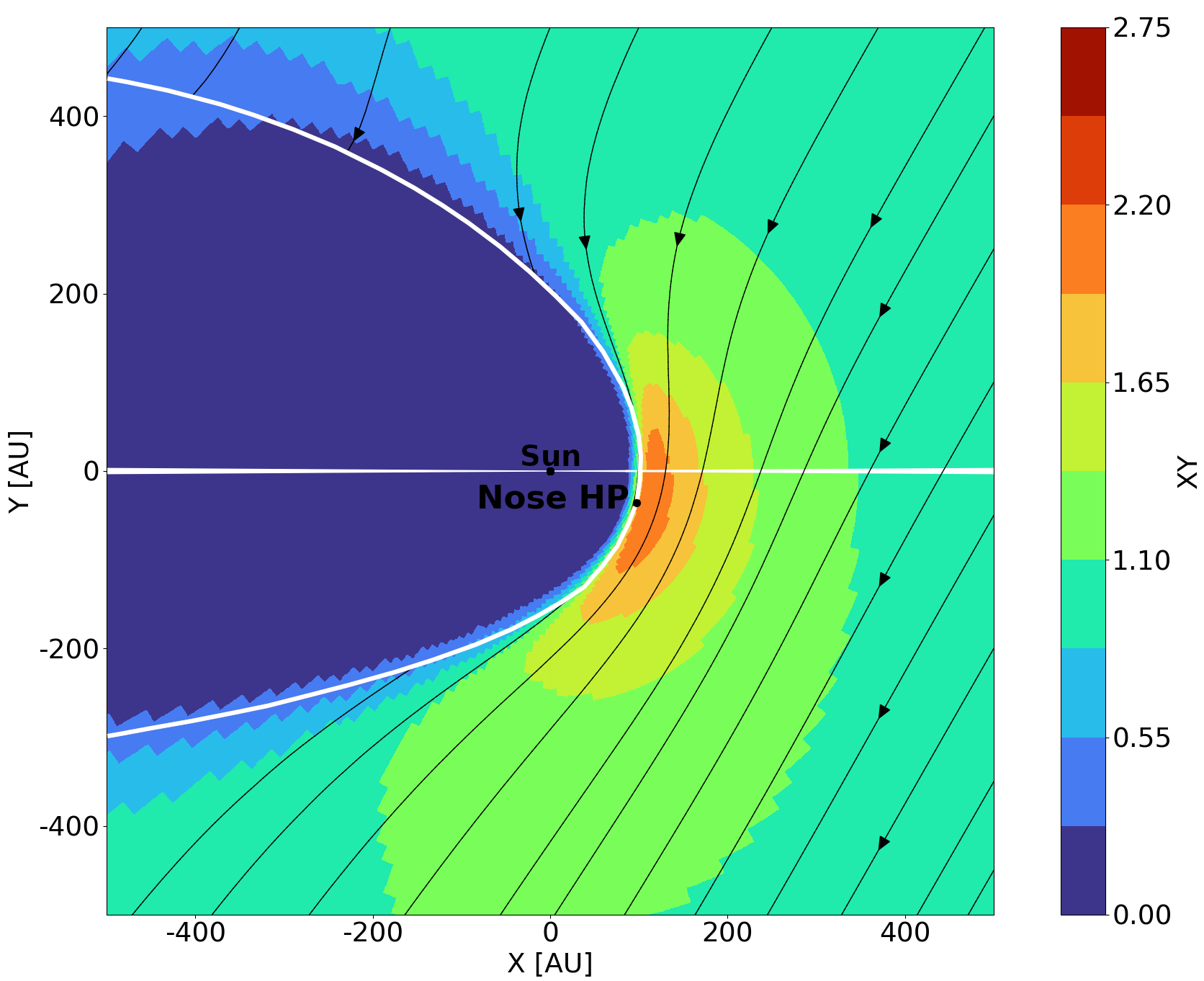}}
            \caption{Iso SW without the IMF (compare to
            Fig.~\ref{fig:fieldlines_60_2}\subref{fig:fieldlines_60_2_iso_no_bsw})}
            \label{fig:fieldlines_60_3_iso_no_bsw}
        \end{subfigure}
        \vskip\baselineskip
        \begin{subfigure}[b]{0.465\textwidth}
            \centering
            \resizebox{\hsize}{!}{\includegraphics{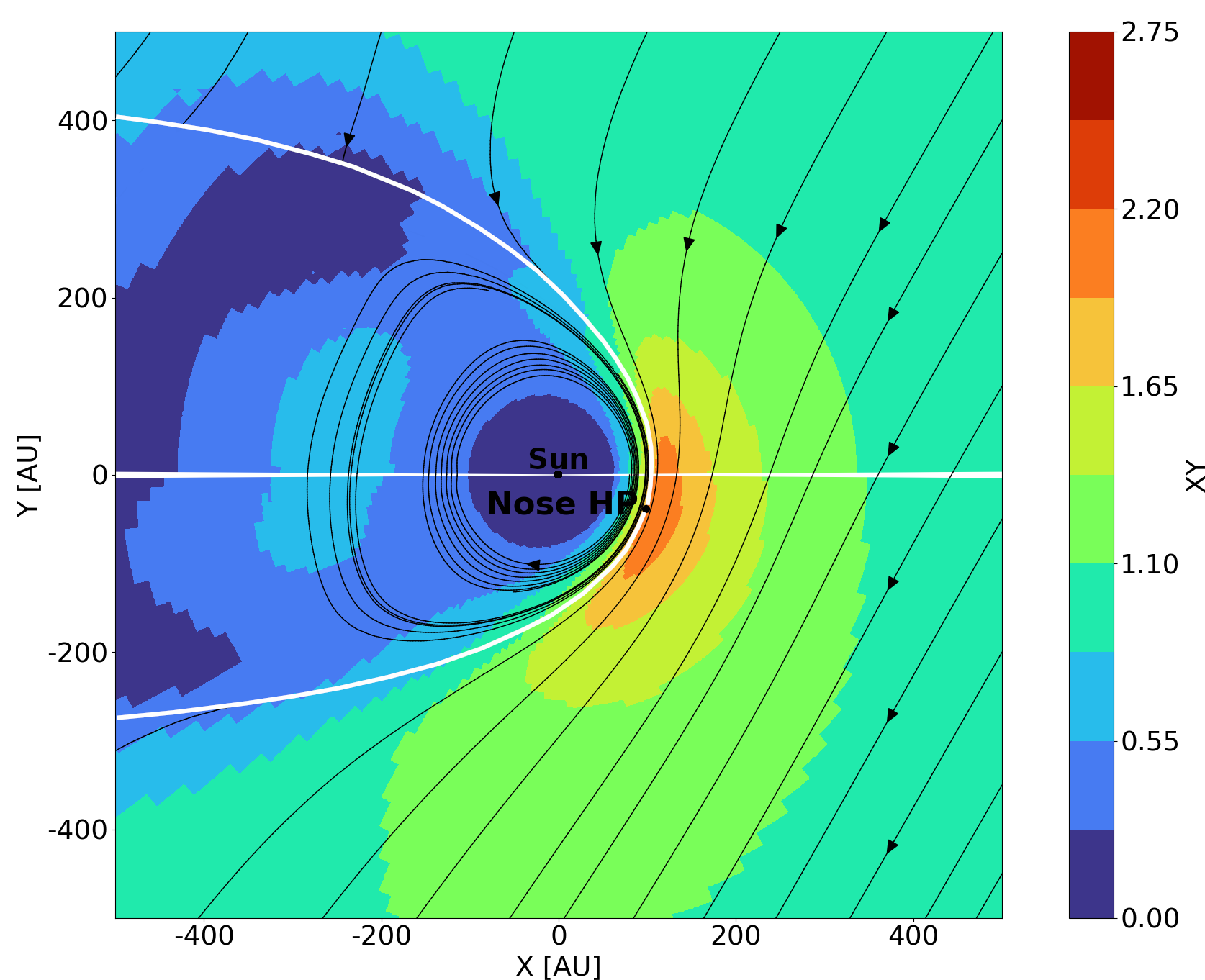}}
            \caption{Iso SW with the IMF (compare to
            Fig.~\ref{fig:fieldlines_60_2}\subref{fig:fieldlines_60_2_iso})}
            \label{fig:fieldlines_60_3_iso}
        \end{subfigure}
        \vskip\baselineskip
        \begin{subfigure}[b]{0.465\textwidth}
            \centering
            \resizebox{\hsize}{!}{\includegraphics{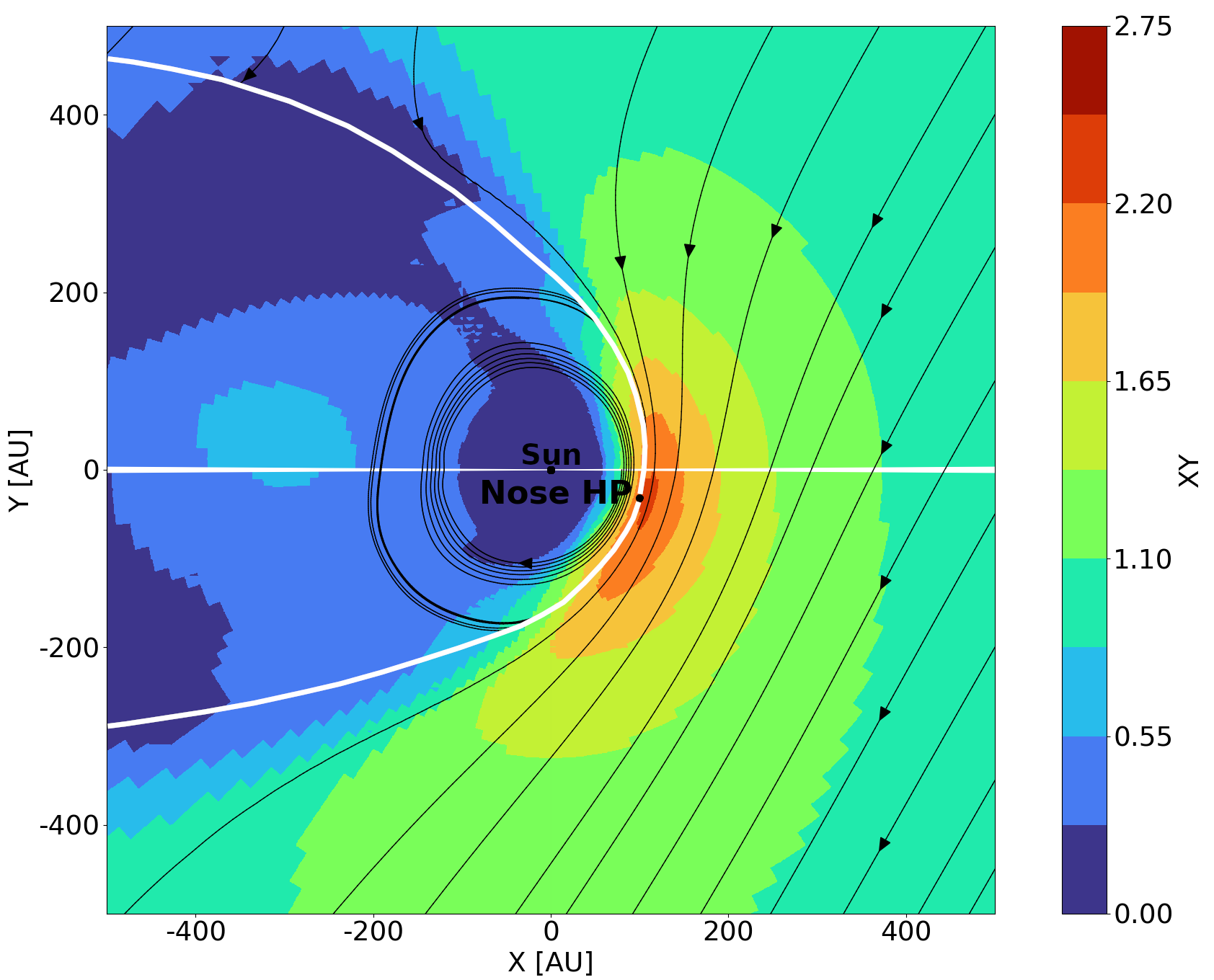}}
            \caption{Anis SW with the IMF (compare to
            Fig.~\ref{fig:fieldlines_60_2}\subref{fig:fieldlines_60_2_anis})}
            \label{fig:fieldlines_60_3_anis}
        \end{subfigure}
        \caption{Magnetic fieldlines (black) shown for the magnetic
        field magnitude for the inclination angle $\alpha = 60\degr$ and
        for ISMF intensity $3 \mu G$}
        \label{fig:fieldlines_60_3}
    \end{figure}

    \begin{figure}
        \centering
        \begin{subfigure}[b]{0.465\textwidth}
            \centering
            \resizebox{\hsize}{!}{\includegraphics{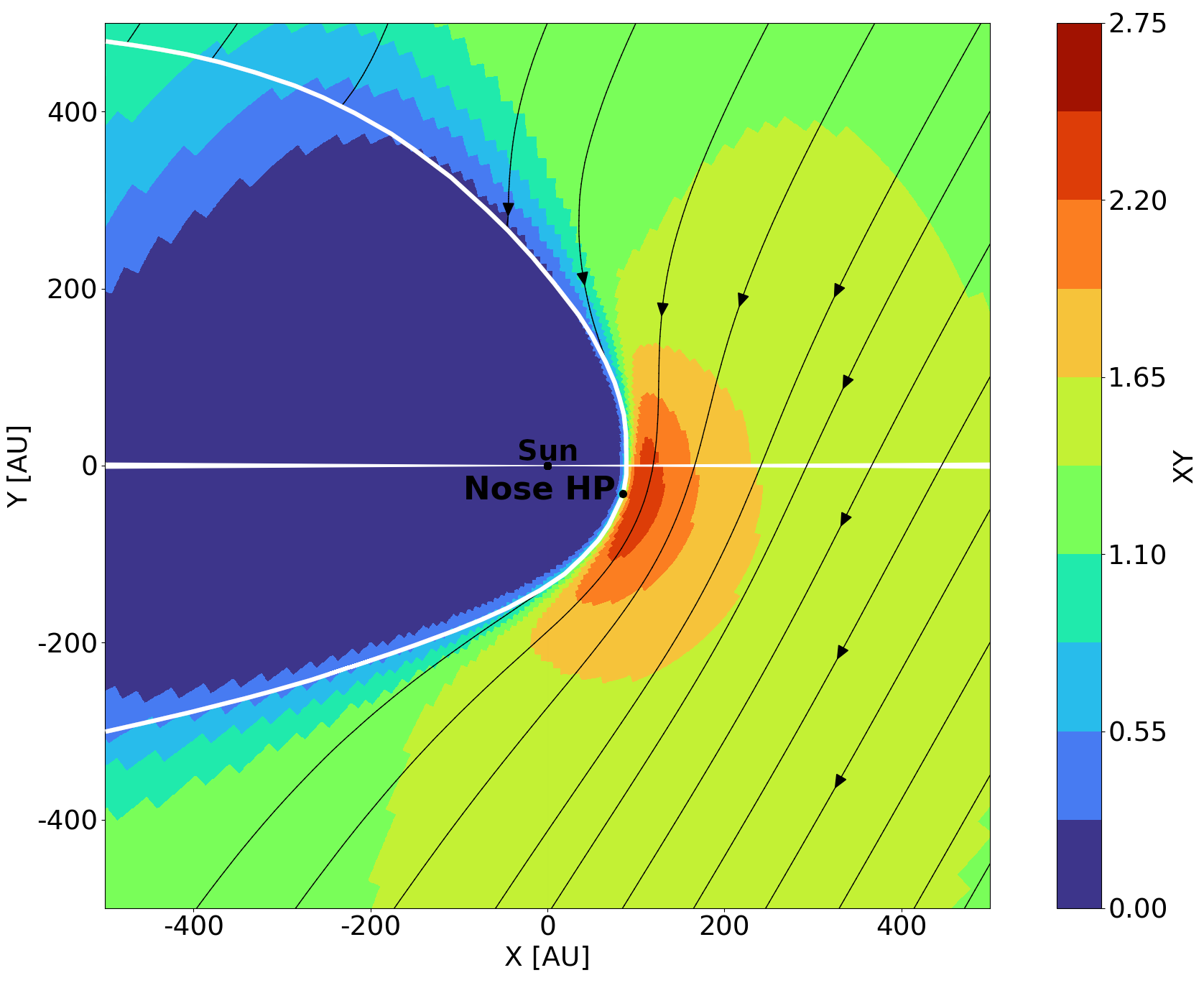}}
            \caption{Iso SW without the IMF (compare
            to Fig.~\ref{fig:fieldlines_60_2}\subref{fig:fieldlines_60_2_iso_no_bsw}
            and Fig.~\ref{fig:fieldlines_60_3}\subref{fig:fieldlines_60_3_iso_no_bsw})}
            \label{fig:fieldlines_60_4_iso_no_bsw}
        \end{subfigure}
        \vskip\baselineskip
        \begin{subfigure}[b]{0.465\textwidth}
            \centering
            \resizebox{\hsize}{!}{\includegraphics{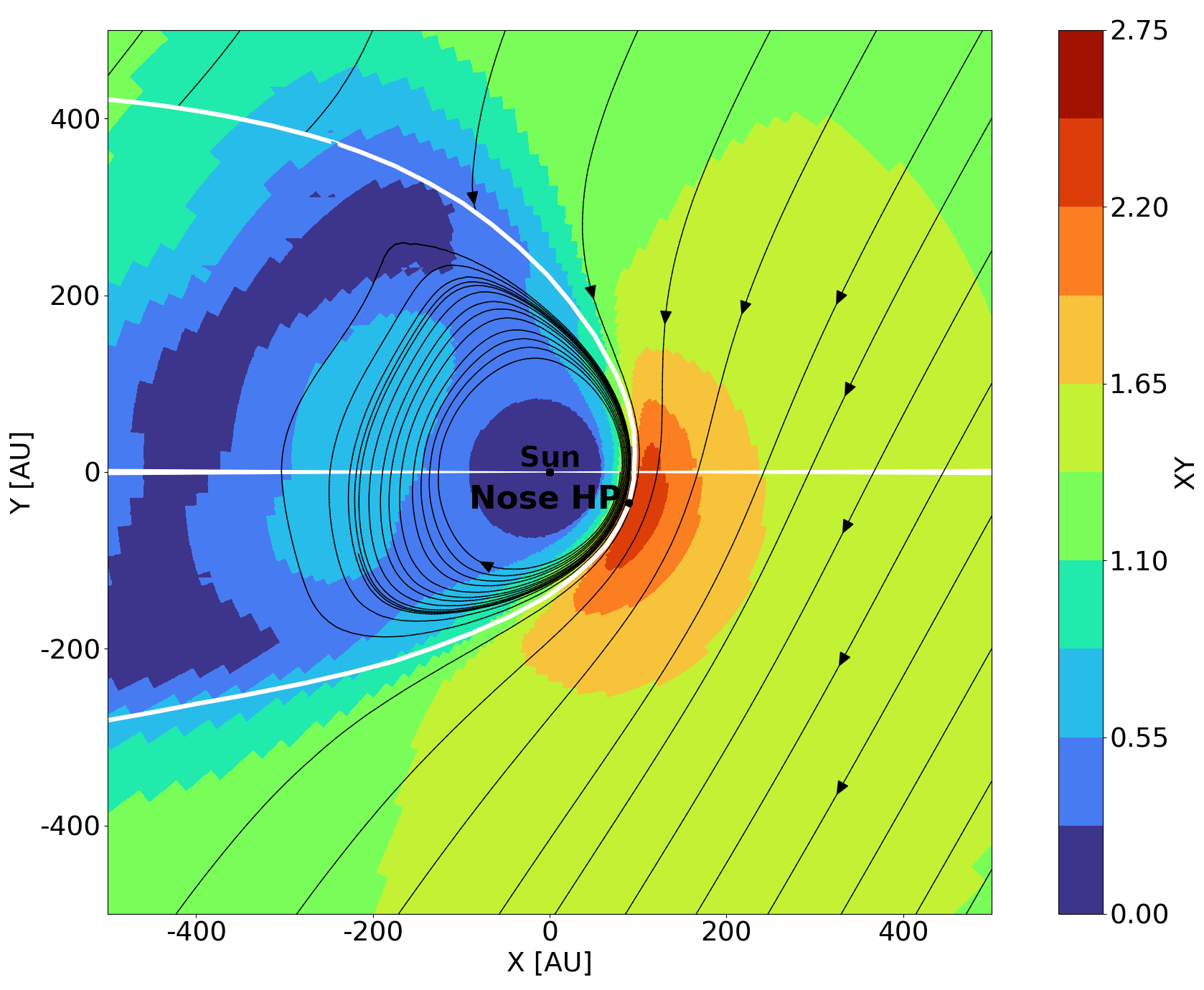}}
            \caption{Iso SW with the IMF (compare
            to Fig.~\ref{fig:fieldlines_60_2}\subref{fig:fieldlines_60_2_iso}
            and Fig.~\ref{fig:fieldlines_60_3}\subref{fig:fieldlines_60_3_iso})}
            \label{fig:fieldlines_60_4_iso}
        \end{subfigure}
        \vskip\baselineskip
        \begin{subfigure}[b]{0.465\textwidth}
            \centering
            \resizebox{\hsize}{!}{\includegraphics{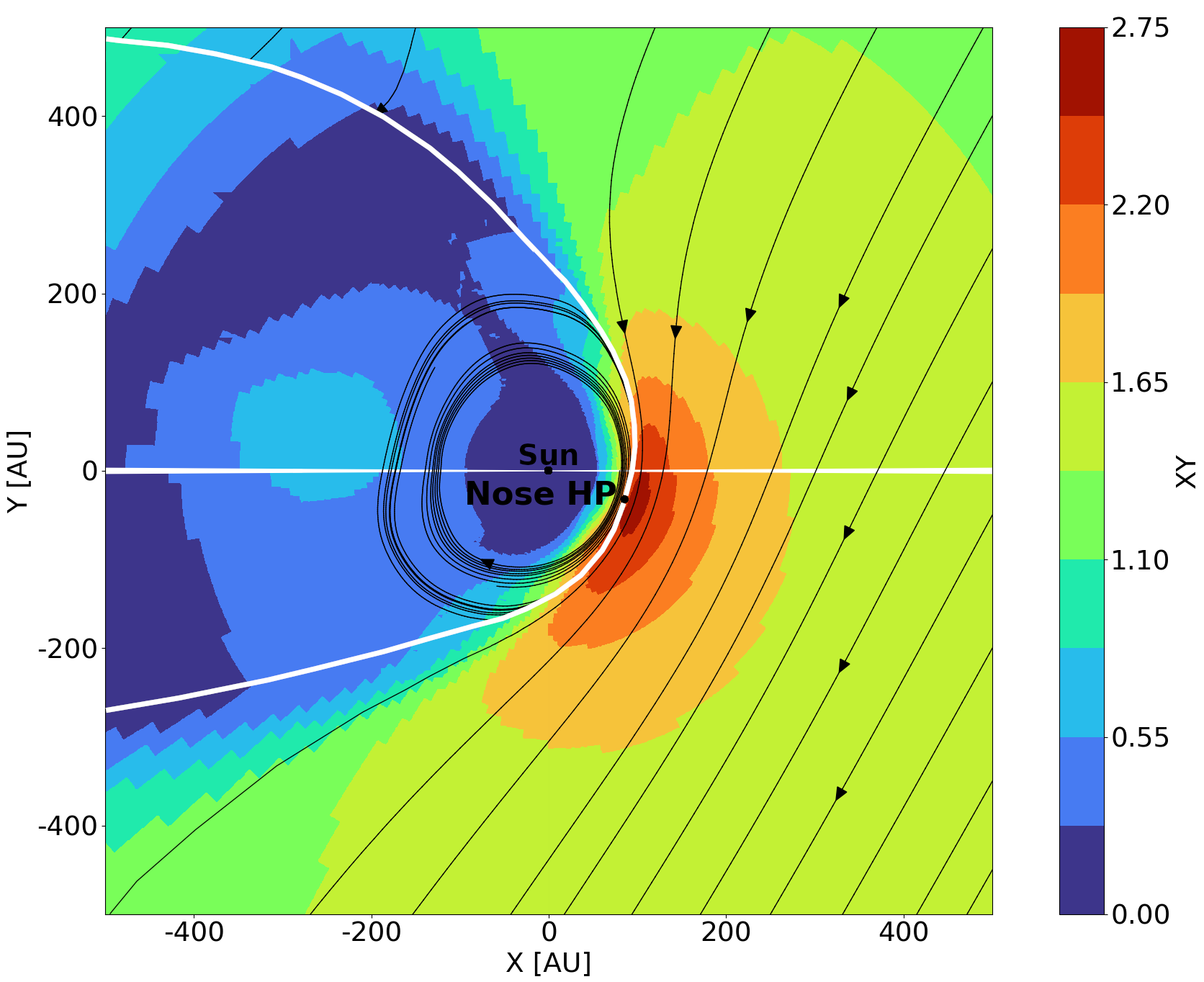}}
            \caption{Anis SW with the IMF (compare
            to Fig.~\ref{fig:fieldlines_60_2}\subref{fig:fieldlines_60_2_anis}
            and Fig.~\ref{fig:fieldlines_60_3}\subref{fig:fieldlines_60_3_anis})}
            \label{fig:fieldlines_60_4_anis}
        \end{subfigure}
        \caption{Magnetic fieldlines (black) shown for the magnetic
        field magnitude for the inclination angle $\alpha = 60\degr$ and
        for ISMF intensity $4 \mu G$}
        \label{fig:fieldlines_60_4}
    \end{figure}

    \begin{figure}
        \centering
        \begin{subfigure}[b]{0.465\textwidth}
            \centering
            \resizebox{\hsize}{!}{\includegraphics{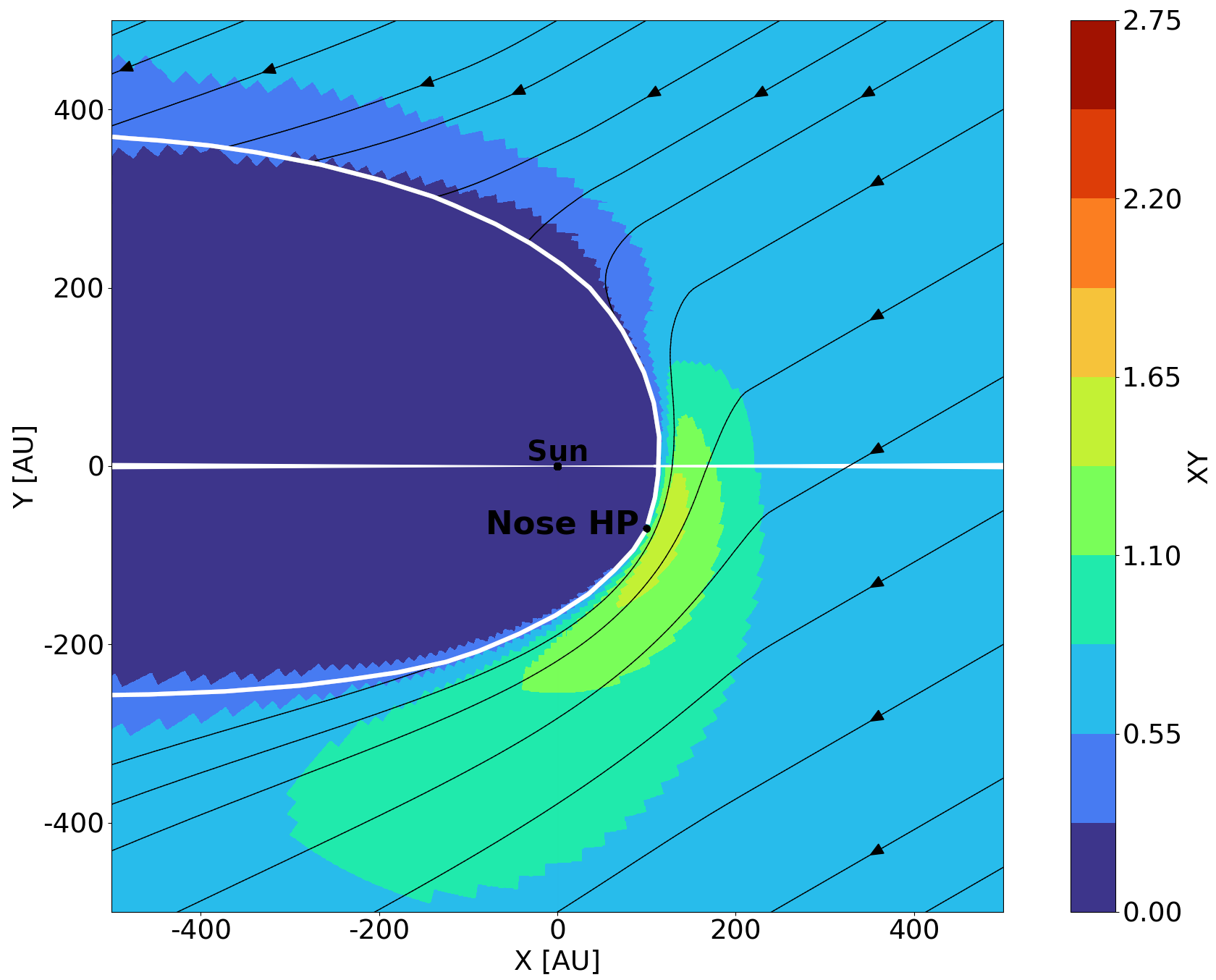}}
            \caption{ISMF intensity $2 \mu G$}
            \label{fig:fieldlines_30_2_iso_no_bsw}
        \end{subfigure}
        \vskip\baselineskip
        \begin{subfigure}[b]{0.465\textwidth}
            \centering
            \resizebox{\hsize}{!}{\includegraphics{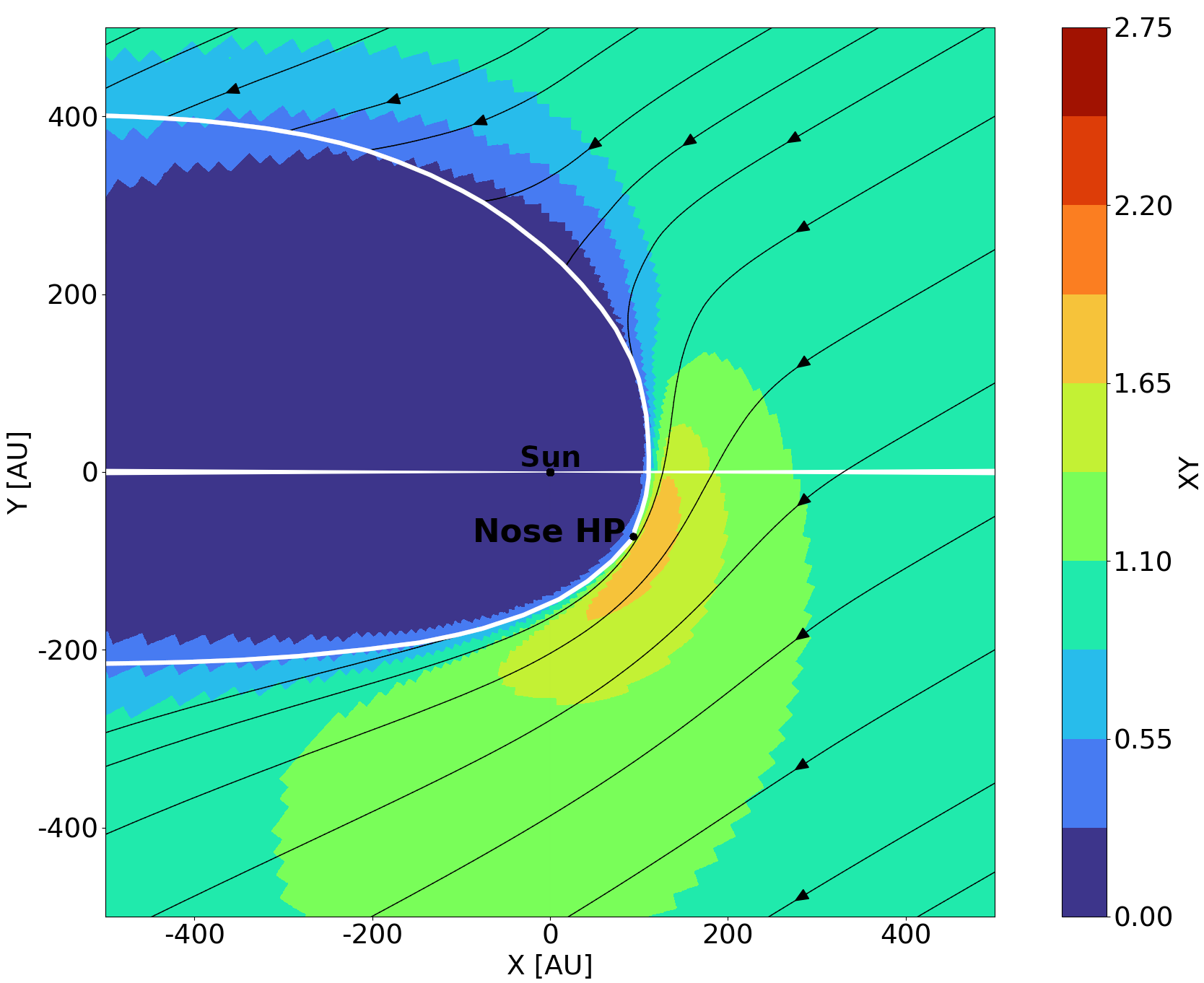}}
            \caption{ISMF intensity $3 \mu G$}
            \label{fig:fieldlines_30_3_iso_no_bsw}
        \end{subfigure}
        \vskip\baselineskip
        \begin{subfigure}[b]{0.465\textwidth}
            \centering
            \resizebox{\hsize}{!}{\includegraphics{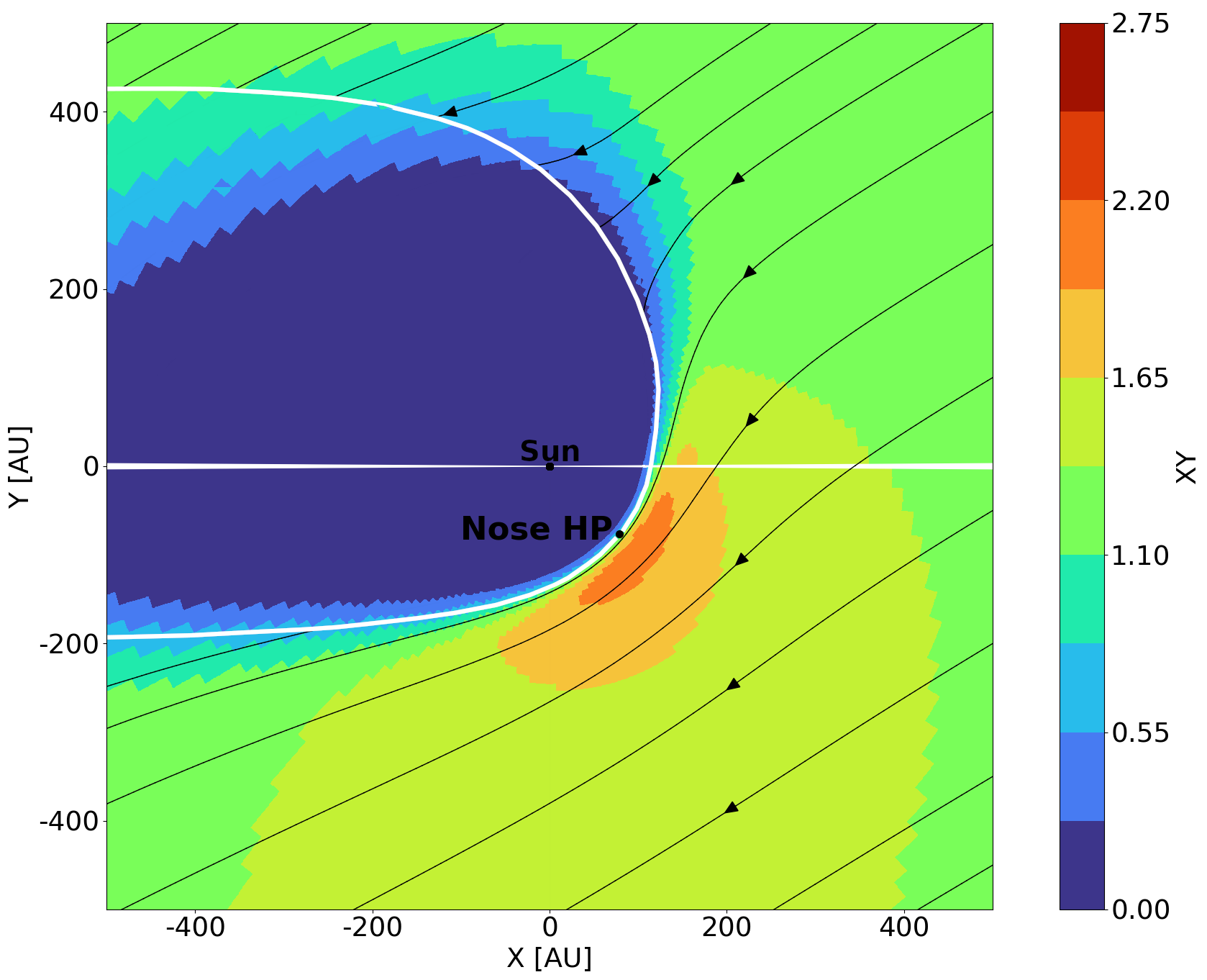}}
            \caption{ISMF intensity $4 \mu G$}
            \label{fig:fieldlines_30_4_iso_no_bsw}
        \end{subfigure}
        \caption{Magnetic fieldlines (black) shown for the magnetic
        field magnitude for the inclination angle $\alpha = 30\degr$, without
        the IMF, and with iso SW}
        \label{fig:fieldlines_30_iso_no_bsw}
    \end{figure}

    The next sequence of figures (
    \ref{fig:fieldlines_60_3}\subref{fig:fieldlines_60_3_iso_no_bsw},
    \ref{fig:fieldlines_60_3}\subref{fig:fieldlines_60_3_iso},
    \ref{fig:fieldlines_60_3}\subref{fig:fieldlines_60_3_anis} and
    \ref{fig:fieldlines_60_4}\subref{fig:fieldlines_60_4_iso_no_bsw},
    \ref{fig:fieldlines_60_4}\subref{fig:fieldlines_60_4_iso},
    \ref{fig:fieldlines_60_4}\subref{fig:fieldlines_60_4_anis})
    describes the same results as Figs.
    \ref{fig:fieldlines_60_2}\subref{fig:fieldlines_60_2_iso_no_bsw},
    \ref{fig:fieldlines_60_2}\subref{fig:fieldlines_60_2_iso}, and
    \ref{fig:fieldlines_60_2}\subref{fig:fieldlines_60_2_anis}
    for an ISMF intensity of $3 \mu G$ and $4 \mu G$ for $\alpha = 60\degr$.
    Figures
    \ref{fig:fieldlines_60_2}\subref{fig:fieldlines_60_2_iso_no_bsw},
    \ref{fig:fieldlines_60_3}\subref{fig:fieldlines_60_3_iso_no_bsw},
    and \ref{fig:fieldlines_60_4}\subref{fig:fieldlines_60_4_iso_no_bsw} (case one)
    show that the greater the ISMF intensity,
    the closer the HP nose is to the Sun and the greater
    the distance of the heliosphere surface from the x-axis towards the tail.
    It is clear that case one, where an IMF is not included,
    is different from case two and case three.
    A comparison of Figs.
    \ref{fig:fieldlines_60_2}\subref{fig:fieldlines_60_2_iso} with
    \ref{fig:fieldlines_60_2}\subref{fig:fieldlines_60_2_anis},
    \ref{fig:fieldlines_60_3}\subref{fig:fieldlines_60_3_iso} with
    \ref{fig:fieldlines_60_3}\subref{fig:fieldlines_60_3_anis},
    and \ref{fig:fieldlines_60_4}\subref{fig:fieldlines_60_4_iso} with
    \ref{fig:fieldlines_60_4}\subref{fig:fieldlines_60_4_anis}
    reveals a different structure of the heliosphere regardless
    of the ISMF intensity.
    A comparison of Figs.
    \ref{fig:fieldlines_60_2}\subref{fig:fieldlines_60_2_iso_no_bsw},
    \ref{fig:fieldlines_60_2}\subref{fig:fieldlines_60_2_iso},
    \ref{fig:fieldlines_60_2}\subref{fig:fieldlines_60_2_anis} with
    \ref{fig:fieldlines_60_3}\subref{fig:fieldlines_60_3_iso_no_bsw},
    \ref{fig:fieldlines_60_3}\subref{fig:fieldlines_60_3_iso},
    \ref{fig:fieldlines_60_3}\subref{fig:fieldlines_60_3_anis} and with
    \ref{fig:fieldlines_60_4}\subref{fig:fieldlines_60_4_iso_no_bsw},
    \ref{fig:fieldlines_60_4}\subref{fig:fieldlines_60_4_iso},
    \ref{fig:fieldlines_60_4}\subref{fig:fieldlines_60_4_anis} shows
    that the greater the ISMF intensity,
    the greater the distance of the heliosphere surface
    from the x-axis towards the tail.
    However, in each case, the nose of the HP clearly
    deviates from the direction of the LISM velocity perpendicular
    to the maximum of the ISMF intensity.

    The next sequence of figures,
    \ref{fig:fieldlines_30_iso_no_bsw}\subref{fig:fieldlines_30_2_iso_no_bsw},
    \ref{fig:fieldlines_30_iso_no_bsw}\subref{fig:fieldlines_30_3_iso_no_bsw},
    and \ref{fig:fieldlines_30_iso_no_bsw}\subref{fig:fieldlines_30_4_iso_no_bsw},
    describes case one (iso SW without the IMF) for the ISMF
    intensity of $2 \mu G$, $3 \mu G$, and $4 \mu G$ for $\alpha = 30\degr$.
    Figures
    \ref{fig:fieldlines_30_iso_no_bsw}\subref{fig:fieldlines_30_2_iso_no_bsw},
    \ref{fig:fieldlines_30_iso_no_bsw}\subref{fig:fieldlines_30_3_iso_no_bsw},
    and \ref{fig:fieldlines_30_iso_no_bsw}\subref{fig:fieldlines_30_4_iso_no_bsw}
    show that the HP nose for the inclination angle $\alpha = 30\degr$
    deviates from the LISM velocity more than for the angle $\alpha= 60\degr$
    (compare Figs.
    \ref{fig:fieldlines_60_2}\subref{fig:fieldlines_60_2_iso_no_bsw},
    \ref{fig:fieldlines_60_3}\subref{fig:fieldlines_60_3_iso_no_bsw},
    and \ref{fig:fieldlines_60_4}\subref{fig:fieldlines_60_4_iso_no_bsw})
    regardless of the ISMF intensity \citep[see][]{fahr1986,fahr1988}.
    Furthermore, the heliosphere shape differs for varies ISMF intensity.
    The next sequence of figures, \ref{fig:fieldlines_30_iso}\subref{fig:fieldlines_30_3_iso} and   \ref{fig:fieldlines_30_iso}\subref{fig:fieldlines_30_4_iso},
    describe case two (isotropic SW with the IMF) for the ISMF
    intensity of $3 \mu G$ and $4 \mu G$ for $\alpha = 30\degr$.
    In case two, Figs.
    \ref{fig:fieldlines_30_iso}\subref{fig:fieldlines_30_3_iso} and
    \ref{fig:fieldlines_30_iso}\subref{fig:fieldlines_30_4_iso},
    with the IMF and the isotropic SW for inclination angle
    $\alpha = 30\degr$ show differences in the heliosphere structures
    for the ISMF intensities of $3 \mu G$ and $4 \mu G$.
    The next sequence of Figs.
    \ref{fig:fieldlines_30_anis}\subref{fig:fieldlines_30_2_anis} and
    \ref{fig:fieldlines_30_anis}\subref{fig:fieldlines_30_4_anis}
    describes case three (SW with the IMF, anis SW)
    for the ISMF intensities of $2 \mu G$ and $4 \mu G$ and the inclination angle of $\alpha = 30\degr$.
    In case three, the Figs.
    \ref{fig:fieldlines_30_anis}\subref{fig:fieldlines_30_2_anis} and
    \ref{fig:fieldlines_30_anis}\subref{fig:fieldlines_30_4_anis},
    with the IMF, and the anisotropic SW for an inclination angle of $\alpha = 30\degr,$
    show even greater differences in the heliosphere structures for
    the ISMF intensities of $2 \mu G$ and $4 \mu G$.

    \section{Summary and conclusions}
    \label{sec:SummaryAndConclusions}

    The three different models of the heliosphere were created
    using the 3D MHD numerical program for three different boundary conditions:
    \begin{enumerate}
        \item the interaction of the isotropic SW (maximum of an 11-year solar cycle)
        with the LISM, without considering the IMF,
        \item the interaction of the isotropic SW with the LISM,
        considering the IMF,
        \item the interaction of the anisotropic SW (minimum of an 11-year solar cycle)
        with the LISM, and considering the IMF.
    \end{enumerate}

    The purpose of this article is to define the nose of the
    HP and investigate the differences in its location
    resulting from a dependence on the intensity and direction of the ISMF,
    which is still not well known, but which has a significant impact on the HP nose movement,
    as we have tried to demonstrate in this paper.
    We explored the differences and similarities between the three models,
    taking into account the different study setups.
    We analysed the ISMF direction parallel, perpendicular,
    and oblique to the LISM velocity vector.

    For the ISMF parallel to the LISM velocity, the heliosphere is axisymmetric.
    What is characteristic in this case is that the more the ISMF compresses the HP,
    the greater the ISMF intensities are.
    The HP nose is in each case at the intersection of the x-axis with the HP.
    Simultaneously, the greater the ISMF intensity,
    the farther the HP nose is from the Sun (see Figs.
    \ref{fig:streamlines}\subref{fig:streamlines2},
    \ref{fig:streamlines}\subref{fig:streamlines3},
    and   \ref{fig:streamlines}\subref{fig:streamlines4}).

    For the ISMF perpendicular to the LISM velocity, the heliosphere loses its axial symmetry.
    In the x-y plane, the nose of the HP, squeezed by the perpendicular
    lines of the ISMF field, approaches the Sun, and the profile of the HP
    towards its tail increases its distance from the Sun. In the x-z plane,
    the HP is compressed along the z-axis.
    In the y-z plane, the HP has the shape of a flattened circle.
    The maximum ISMF intensity is in the direction perpendicular to the Sun's
    line of sight (see our Figs.
    \ref{fig:streamlines_fieldlines_4_90}\subref{fig:streamlines_fieldlines_4_90_xy},
    \ref{fig:streamlines_fieldlines_4_90}\subref{fig:streamlines_4_90_xz},
    \ref{fig:streamlines_fieldlines_4_90}\subref{fig:streamlines_fieldlines_4_90_yz} and
    \ref{fig:streamlines_fieldlines_2_90_xy} and compare with
    \citet[][]{ratkiewicz2000}).

    \begin{figure}
        \centering
        \begin{subfigure}[b]{0.465\textwidth}
            \centering
            \resizebox{\hsize}{!}{\includegraphics{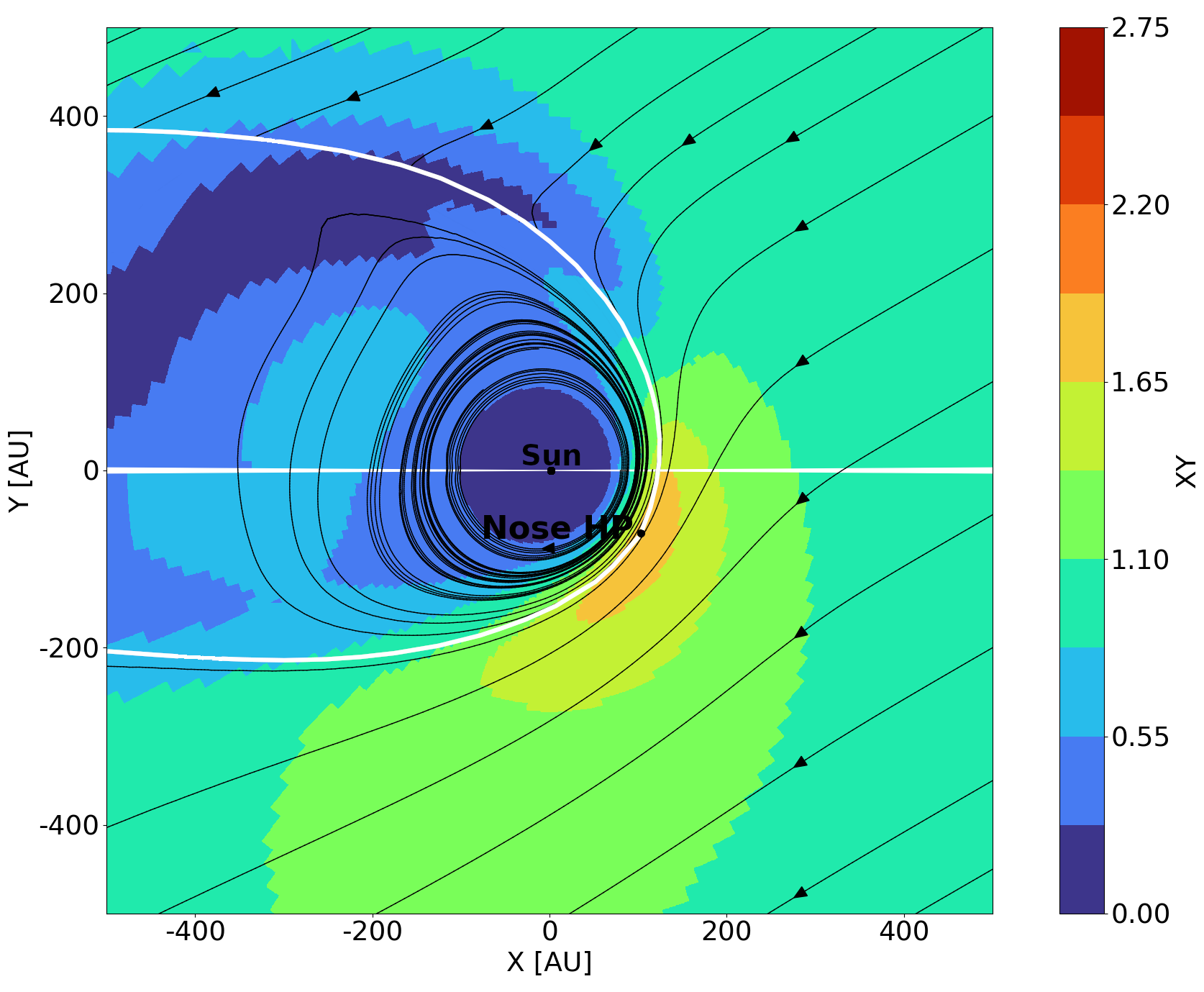}}
            \caption{ISMF intensity $3 \mu G$}
            \label{fig:fieldlines_30_3_iso}
        \end{subfigure}
        \vskip\baselineskip
        \begin{subfigure}[b]{0.465\textwidth}
            \centering
            \resizebox{\hsize}{!}{\includegraphics{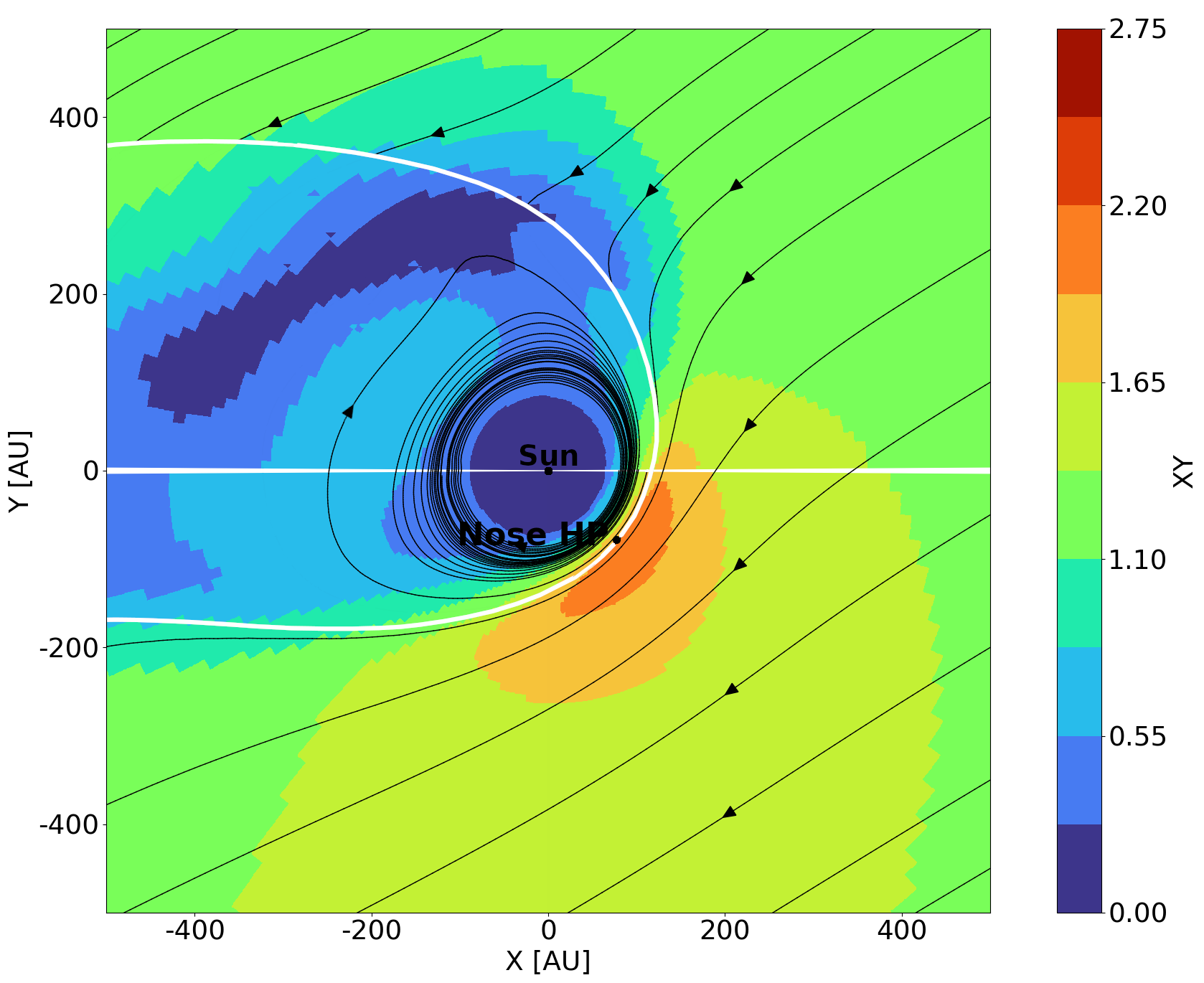}}
            \caption{ISMF intensity $4 \mu G$}
            \label{fig:fieldlines_30_4_iso}
        \end{subfigure}
        \caption{Magnetic fieldlines (black) shown for the magnetic
        field magnitude for the inclination angle $\alpha = 30\degr$, with
        the IMF, and with iso SW}
        \label{fig:fieldlines_30_iso}
    \end{figure}

    \begin{figure}
        \centering
        \begin{subfigure}[b]{0.465\textwidth}
            \centering
            \resizebox{\hsize}{!}{\includegraphics{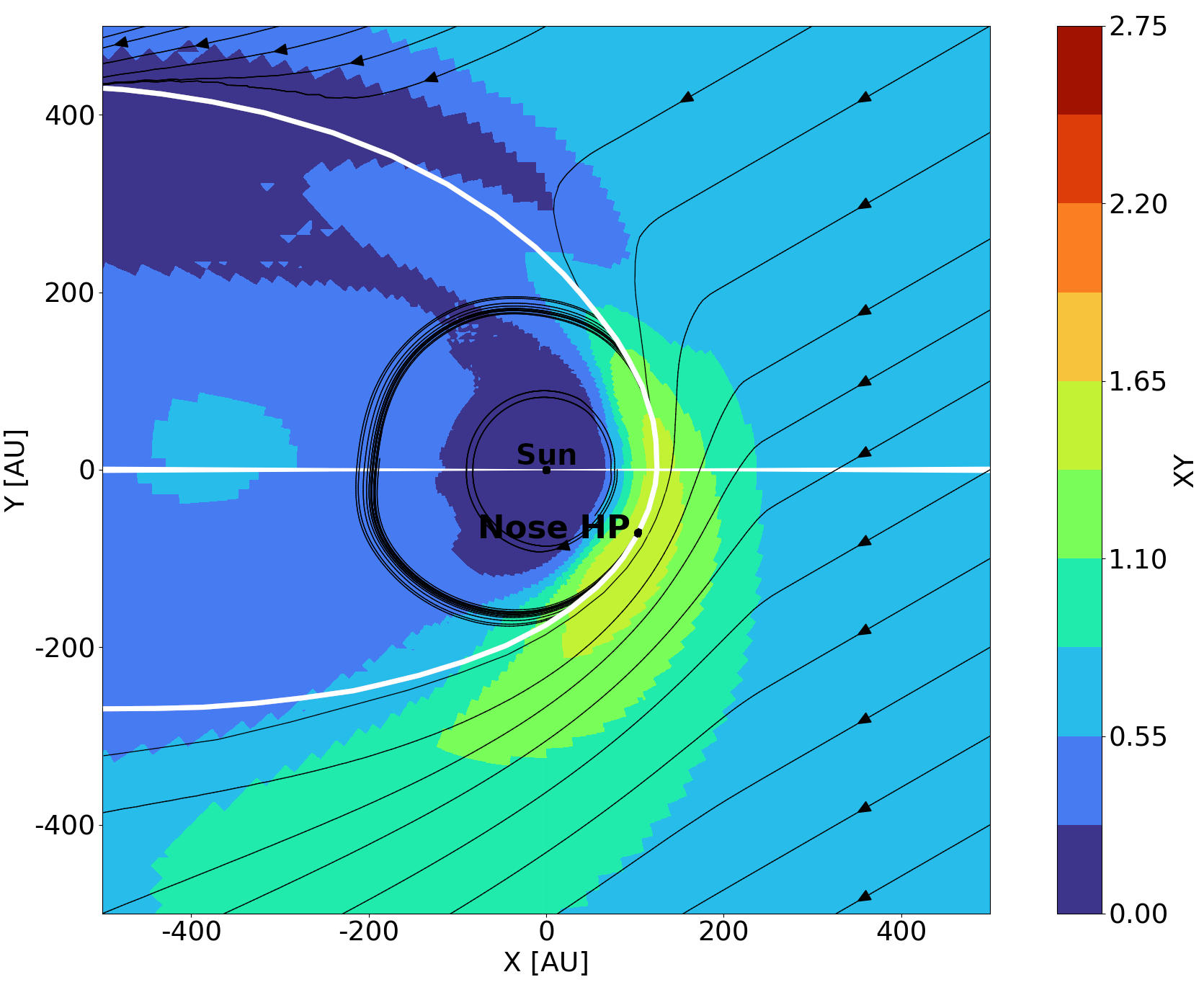}}
            \caption{ISMF intensity $2 \mu G$}
            \label{fig:fieldlines_30_2_anis}
        \end{subfigure}
        \vskip\baselineskip
        \begin{subfigure}[b]{0.465\textwidth}
            \centering
            \resizebox{\hsize}{!}{\includegraphics{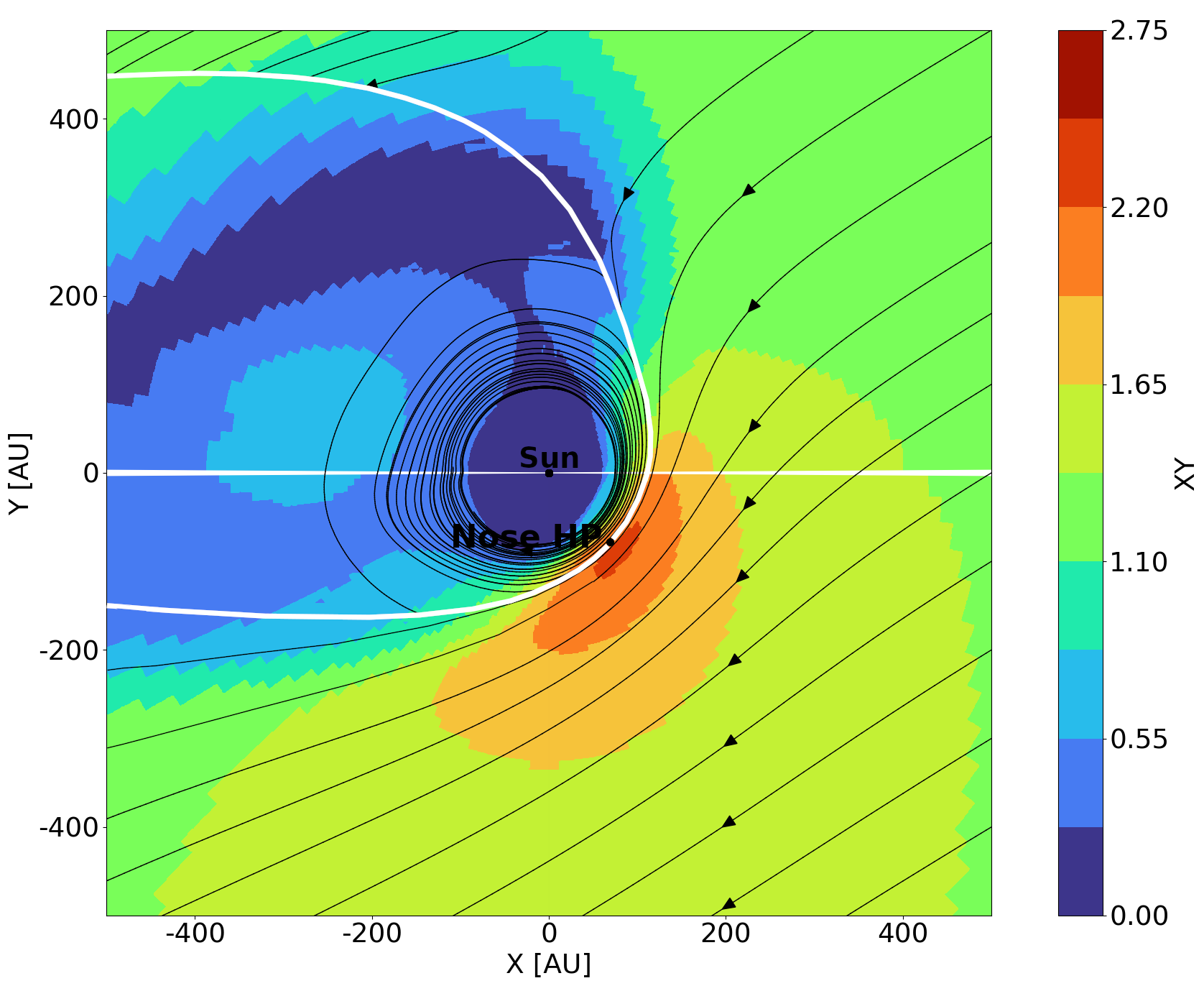}}
            \caption{ISMF intensity $4 \mu G$}
            \label{fig:fieldlines_30_4_anis}
        \end{subfigure}
        \caption{Magnetic fieldlines (black) shown for the magnetic
        field magnitude for the inclination angle $\alpha = 30\degr$, with
        the IMF, and with anis SW}
        \label{fig:fieldlines_30_anis}
    \end{figure}

    In the case of the oblique ISMF direction to the LISM velocity,
    when $\alpha = 60\degr,$ the HP for solar maximum without an IMF
    (case one; see Figs.
    \ref{fig:streamlines_60_2}\subref{fig:streamlines_60_2_iso_no_bsw} and
    \ref{fig:fieldlines_60_2}\subref{fig:fieldlines_60_2_iso_no_bsw})
    is greater than for case two
    with the IMF (see Fig.
    \ref{fig:streamlines_60_2}\subref{fig:streamlines_60_2_iso} and
    \ref{fig:fieldlines_60_2}\subref{fig:fieldlines_60_2_iso}).
    The heliosphere for the minimum 11-year cycle of solar activity with the
    IMF taken into account (case three, see Figs.
    \ref{fig:streamlines_60_2}\subref{fig:streamlines_60_2_anis} and
    \ref{fig:fieldlines_60_2}\subref{fig:fieldlines_60_2_anis})
    differs from the previous cases (Figs.
    \ref{fig:streamlines_60_2}\subref{fig:streamlines_60_2_iso} and
    \ref{fig:fieldlines_60_2}\subref{fig:fieldlines_60_2_iso}).
    In all three cases, the nose of the HP clearly deviates
    from the direction of the LISM velocity.

    A comparison of Figs.
    \ref{fig:fieldlines_60_2}\subref{fig:fieldlines_60_2_iso_no_bsw},
    \ref{fig:fieldlines_60_3}\subref{fig:fieldlines_60_3_iso_no_bsw},
    and \ref{fig:fieldlines_60_4}\subref{fig:fieldlines_60_4_iso_no_bsw}
    (case one) clearly shows that the
    greater the ISMF intensity, the closer the HP nose is to the
    Sun and the greater the distance of the heliosphere surface
    from the x-axis towards the tail.
    Figures
    \ref{fig:fieldlines_30_iso_no_bsw}\subref{fig:fieldlines_30_2_iso_no_bsw},
    \ref{fig:fieldlines_30_iso_no_bsw}\subref{fig:fieldlines_30_3_iso_no_bsw},
    and \ref{fig:fieldlines_30_iso_no_bsw}\subref{fig:fieldlines_30_4_iso_no_bsw}
    show that the HP nose for the inclination angle $\alpha = 30\degr$
    deviates from the LISM velocity more than for angle $\alpha = 60\degr,$
    regardless of the ISMF intensity \citep[see][]{fahr1986, fahr1988}.
    Besides this, the heliosphere shape differs for various ISMF intensities.

    The following considerations concern only cases two and three,
    in which the IMF is included.
    Comparisons of Figs.
    \ref{fig:fieldlines_60_2}\subref{fig:fieldlines_60_2_iso} with
    \ref{fig:fieldlines_60_2}\subref{fig:fieldlines_60_2_anis},
    \ref{fig:fieldlines_60_3}\subref{fig:fieldlines_60_3_iso} with
    \ref{fig:fieldlines_60_3}\subref{fig:fieldlines_60_3_anis},
    and \ref{fig:fieldlines_60_4}\subref{fig:fieldlines_60_4_iso} with
    \ref{fig:fieldlines_60_4}\subref{fig:fieldlines_60_4_anis}
    show a different structure of the heliosphere regardless of the ISMF intensity.
    Comparisons of Figs.
    \ref{fig:fieldlines_60_2}\subref{fig:fieldlines_60_2_iso_no_bsw},
    \ref{fig:fieldlines_60_2}\subref{fig:fieldlines_60_2_iso},
    \ref{fig:fieldlines_60_2}\subref{fig:fieldlines_60_2_anis},
    with
    \ref{fig:fieldlines_60_3}\subref{fig:fieldlines_60_3_iso_no_bsw},
    \ref{fig:fieldlines_60_3}\subref{fig:fieldlines_60_3_iso},
    \ref{fig:fieldlines_60_3}\subref{fig:fieldlines_60_3_anis},
    and with
    \ref{fig:fieldlines_60_4}\subref{fig:fieldlines_60_4_iso_no_bsw},
    \ref{fig:fieldlines_60_4}\subref{fig:fieldlines_60_4_iso},
    \ref{fig:fieldlines_60_4}\subref{fig:fieldlines_60_4_anis}
    show that the greater the ISMF intensity,
    the greater the distance of the heliosphere surface
    from the x-axis towards the tail.
    In each case, the nose of the HP clearly deviates from
    the direction of the LISM velocity perpendicular
    to the maximum of the ISMF intensity.

    Figures
    \ref{fig:fieldlines_30_iso}\subref{fig:fieldlines_30_3_iso} and
    \ref{fig:fieldlines_30_iso}\subref{fig:fieldlines_30_4_iso},
    with the IMF, and with isotropic SW (case two),
    for an inclination angle of  $\alpha = 30\degr,$ show different structures
    of the heliosphere for the ISMF intensities of $3 \mu G$ and $4 \mu G$.
    Figures
    \ref{fig:fieldlines_30_anis}\subref{fig:fieldlines_30_2_anis} and
    \ref{fig:fieldlines_30_anis}\subref{fig:fieldlines_30_4_anis},
    with the IMF, and with anisotropic SW (case three)
    for the ISMF intensities of  $2 \mu G$ and $4 \mu G$ and an inclination angle of $\alpha = 30\degr$
    present even more different structures of the heliosphere.

    The analysis carried out in this article for the three
    simplest cases of heliosphere models leads to the conclusion
    that the HP nose defined in the works of
    \citet[see][]{fahr1986, fahr1988} and \citet[][]{ratkiewiczbaraniecka2023},
    and this article, is always located in the direction perpendicular
    to the maximum intensity of the ISMF.
    The displacement of the HP nose depends upon the direction and intensity of the ISMF,
    with the structure of the heliosphere and the shape
    of the HP, depending on the 11-year cycle of solar activity.
    The discussion on the nose of the HP also revealed the richness of
    the structures and shapes of the heliosphere.

    In the context of the planned space mission to send the IP to a distance of $1000 AU$ from the Sun for the first time
    in human history \citep{brandt2022},
    our study may shed light on the question of which direction to send the IP.
    Based on these results, the conclusion arises that further research
    should continue through the introduction of issues such as current sheet,
    reconnection, cosmic rays, instability, or turbulence
    into the models (\citet[][]{kornbleuth2021a, mccomasschwadron2006, mccomas2008, opher2017,
    opher2021, opher2023} and \citet[][]{strumikratkiewicz2022}).

    We emphasise, however, that the assumptions about the constant velocity, temperature,
    intensity and direction of the magnetic field, and the density of plasma and neutral atoms
    adopted so far as boundary conditions in LISM are accepted,
    but we do not know what they are for sure \citep{strumikratkiewicz2022}.
    The results from all MHD heliosphere models are far from realistic,
    since the MHD approach has significant simplifications.
    In our opinion, taking into account the above statements,
    as well as the phenomena and processes mentioned earlier,
    and the fact that the interstellar field can be structured and heterogeneous,
    the task to be solved requires more sophisticated numerical
    programs based on the use of artificial intelligence techniques that we
    intend to apply in our future work.

    \section{Acknowledgments}
    \label{sec:Acknowledgments}

    The authors thank Marek Strumik for his support with numerical calculations.
    Calculations were carried out at the Centre of Informatics,
    Tricity Academic Supercomputer \& Network.
    The authors also thank Hans Fahr for reviewing the article and for his
    constructive remarks and suggestions, which increased its quality.

\end{document}